\documentclass[%
 reprint,
 superscriptaddress,
 amsmath,amssymb,
 aps
]{revtex4-1}

\usepackage[dvipdfmx]{graphicx}%
\usepackage{dcolumn}%
\usepackage{bm}%
\usepackage{color}%
\usepackage{siunitx}
\usepackage{here}
\usepackage{mathtools}
\usepackage{nomencl} 
\makenomenclature

\preprint{APS/123-QED}

\usepackage{etoolbox}

\newcommand{\red}[1]{\textcolor{black}{#1}}

\begin{document}

\title{Accurate determination of thermoelectric figure of merit using ac Harman method 
with a four-probe configuration}

\author{K. Okawa}
\email{okawa.k@aist.go.jp}
\affiliation{%
National Metrology Institute of Japan (NMIJ), National Institute of Advanced Industrial Science and Technology (AIST), AIST Central 3, 1-1-1 Umezono, Tsukuba 305-8563, Japan}%
\author{Y. Amagai}%
\affiliation{%
National Metrology Institute of Japan (NMIJ), National Institute of Advanced Industrial Science and Technology (AIST), AIST Central 3, 1-1-1 Umezono, Tsukuba 305-8563, Japan}%
\affiliation{%
Global Research and Development Center for Business by Quantum-AI Technology (G-QuAT), National Institute of Advanced Industrial Science and Technology (AIST), AIST Central 1, 1-1-1 Umezono, Tsukuba 305-8560, Japan}%
\author{N. Sakamoto}%
\affiliation{%
National Metrology Institute of Japan (NMIJ), National Institute of Advanced Industrial Science and Technology (AIST), AIST Central 3, 1-1-1 Umezono, Tsukuba 305-8563, Japan}%
\author{N.-H. Kaneko}%
\affiliation{%
National Metrology Institute of Japan (NMIJ), National Institute of Advanced Industrial Science and Technology (AIST), AIST Central 3, 1-1-1 Umezono, Tsukuba 305-8563, Japan}%
\affiliation{%
Global Research and Development Center for Business by Quantum-AI Technology (G-QuAT), National Institute of Advanced Industrial Science and Technology (AIST), AIST Central 1, 1-1-1 Umezono, Tsukuba 305-8560, Japan}%
\
\begin{abstract}
The ac Harman method has been used for the direct estimation of dimensionless thermoelectric figure of merit ({\ensuremath{zT}}) through ac/dc resistance measurements. However, accurate {\ensuremath{zT}} estimation with a four-probe configuration is difficult owing to the occurrence of a thermal phase-delay in the heat flow with a low frequency current. This study reports an exact solution for {\ensuremath{zT}} estimation by solving the heat conduction equation. The analysis can explain the reverse heat flow, which is the main source of the error in the four-probe configuration, and the experimentally obtained behavior of the frequency dependence of {\ensuremath{zT}} of (Bi,Sb)$_2$Te$_3$. Approximately \SI{20}{\%} of the error is caused by a thermal phase-delay, unless an appropriate current frequency and voltage-terminal position are chosen. Thus, an accurate {\ensuremath{zT}} evaluation using a four-probe configuration at any voltage terminal position is achieved. These findings can lead to interesting thermoelectric metrology and could serve as a powerful tool to search for promising thermoelectric materials.
\end{abstract}

\maketitle


\section{Introduction}
\label{introduction}
Thermoelectric (TE) materials are expected to contribute to power generation in situ as they facilitate the efficient conversion of waste heat into electrical energy \cite{Bell1,Zhang2,He3}. The conversion efficiency of TE materials depends on the dimensionless figure of merit $zT$, which is defined as $zT \equiv S^{2}T/(\rho \kappa)$, where $S$, $T$, $\rho$, and $\kappa$ denote the Seebeck coefficient, absolute temperature, electrical resistivity, and thermal conductivity, respectively \cite{Rowe4}. The accurate evaluation of the $zT$ of TE materials requires independent measurements of the three physical parameters ($S$, $\rho$, and $\kappa$). Generally, such measurements are performed by employing different experimental setups and sample shapes, which can affect the accuracy of $zT$ estimation. In a recent international round-robin test, the interlaboratory uncertainty for $zT$ was estimated to be approximately \SI{10}{\%} to \SI{20}{\%} \cite{Wang5,Wang6,AllenoCoNiSb,HeremansNM}. Consequently, the establishment of standardized and accurate measurement techniques to realize the precise estimation of TE properties is desirable.

The ac Harman method is an alternative method to determine $zT$. It enables the direct determination of $zT$ through ac and dc resistance measurements using the following formula
\begin{eqnarray}
zT = \frac{R_{\rm dc}-R_{\rm ac}}{R_{\rm ac}},
\label{eq1}
\end{eqnarray}
where $R_{\rm dc}$ and $R_{\rm ac}$ denote the resistances of the sample measured via application of dc and ac currents, respectively \cite{Harman7}. The ac Harman method exploits the Peltier effect, which occurs at the sample edges upon the application of the current $I$ to the sample. The ac Harman method has been extensively used for a simple $zT$ estimation procedure in the case of various targets, such as minute crystals, thin films, and module composite structures, because it can be performed based on a simple two-probe configuration \cite{Satake8,Kobayashi9,Kobayashi10,Singh11,Venkatasubramanlan12,Iwasaki13,Korzhuev47,Barako48,Muto49}.

\begin{figure}[]
\centering
\includegraphics[width=8cm,clip]{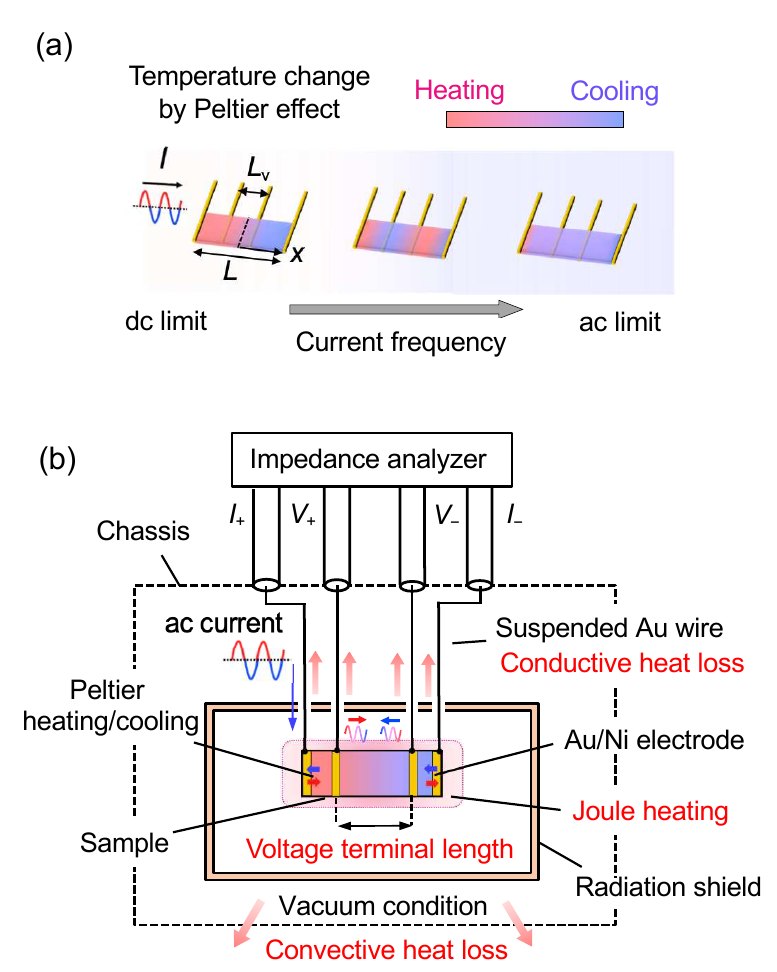}
\caption{(a) Schematic of temperature change by the Peltier effect in a rectangular sample. Through the application of a current $I$, Peltier heating and cooling is produced at the edge of the sample. The sample length and voltage-terminal distance are  given by $L$ and $L_{\rm V}$, respectively. The temperature distribution caused by the Peltier effect varies with the current frequency $f$. (b) Schematic of the experimental setup for the ac Harman method with  the four-probe configuration used in this study. The main sources of error in the measurement are indicated in red text.}
\label{fig1}
\end{figure}

To accurately perform resistance measurements, it is generally more effective to use a four-probe configuration rather than the two-probe configuration. The advantage of the four-probe configuration is that it can discard the influence of wiring (contact) resistance. However, important error factors that are unique to a four-probe configuration exist, which are caused by an inhomogeneous temperature gradient in a particular low-frequency region (Fig. \ref{fig1}(a)). The experimentally estimated $zT$ using a four-probe configuration varies significantly as a function of the distances between the two voltage-terminals $L_{\rm V}$ \cite{Iwasaki14}. Although this error factor is known phenomenologically as a thermal phase delay when using the ac Harman method with the four-probe configuration \cite{Iwasaki15,Downey16}, analytical and experimental validation are lacking.

This study reports a general expression for the ac Harman method with a four-probe configuration. The exact solution of the temperature distribution in the sample was derived from the heat conduction equation that incorporates the Joule effect. Moreover, experiments were performed to estimate $zT$ using Bi-Te materials, and the results are consistent with the exact solution over a wide frequency range. Thus, the proposed analysis facilitates an accurate $zT$ estimation using the ac Harman method with a four-probe configuration at an arbitrary voltage terminal distance. 

\begin{figure}[]
\centering
\includegraphics[width=8cm,clip]{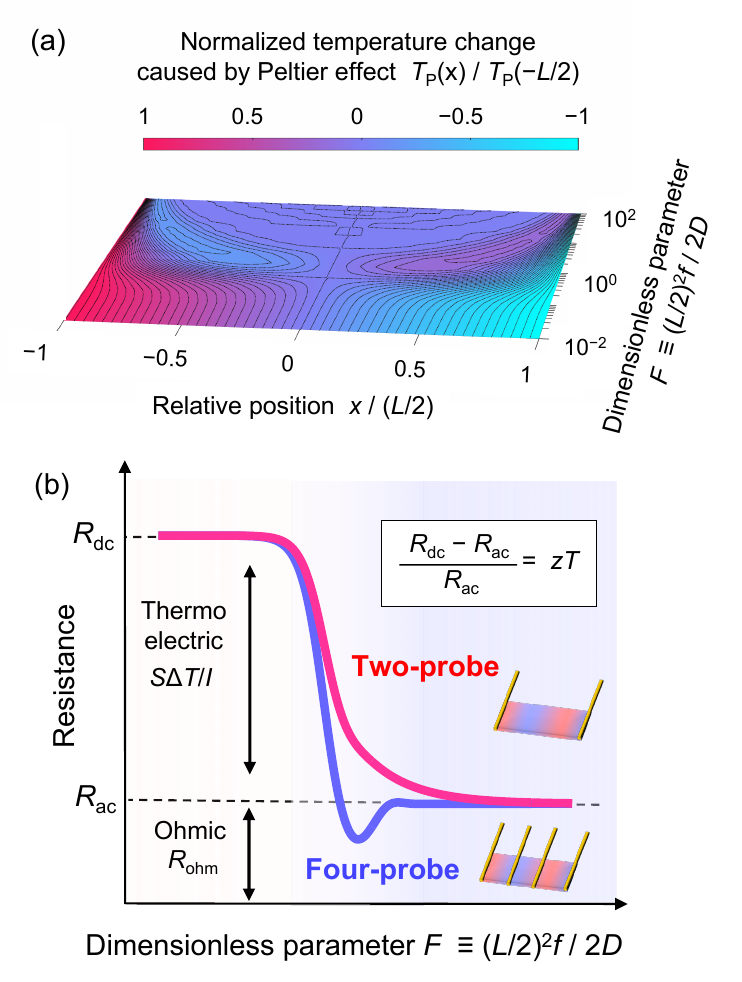}
\caption{(a) Contour plot of the temperature change normalized at one of the sample edges in the relative position $x/(L/2)$ and the dimensionless parameter $F \equiv (L/2)^{2}f/2\red{D}$ plane at the time of the applied root-mean-square ac current, where $\red{D}$  denotes the thermal diffusivity of the sample. (b) Schematic of the measured resistance with two and four-probe configuration as a function of $F$. In the ac measurement, an ohmic resistance, $R_{\rm ohm}$, is observed, while a thermoelectric voltage $S\Delta T/I$ is added in the dc measurement.}
\label{fig2}
\end{figure}

\section{Theory}
\subsection{Exact solution of temperature distribution in a material}
Figure \ref{fig1}(b) shows a schematic of the setup of the ac Harman method with a four-probe configuration. Herein, the case where resistance measurement is performed on a rectangular parallelepipedal sample with a sample length $L$ and cross-sectional area $A$ using the four-probe configuration at a voltage terminal distance $L_{\rm V}$ is considered. The temperature distribution in the sample $T$($x$,$t$) when an ac current with a current density $J(t) = J_{0}\sin\omega t$ can be obtained by solving the following one-dimensional unsteady heat conduction equation,
\begin{eqnarray}
\frac{\partial^2 T(x,t)}{\partial x^2}=\frac{1}{\red{D}}\frac{\partial T(x,t)}{\partial t}-\frac{\rho (J_{0}\sin{\omega t})^{2}}{\kappa},
\label{eq2}
\end{eqnarray}

\noindent
where, $x$, $t$, $J_{0}$, $\omega$, and $\red{D}$ represent the position, time, peak current density, current angular velocity, and thermal diffusivity, respectively. The second term on the right-hand side represents the Joule effect. The Joule heat $\rho J^{2}$ was considered  to incorporate the current amplitude dependence. It was assumed that the values of the physical properties were independent of the temperature.  To simplify the boundary condition, it was assumed that only the Peltier heat $SJT_{0}$ occurred at both ends of the sample $(x = \pm L/2)$, where $T_{0}$ denotes the average temperature of the sample. Consequently, the boundary condition is described as
\begin{eqnarray}
\left.\frac{\partial T(x,t)}{\partial x}\right|_{x=\pm L/2} = \frac{ST_{0}J_{0}}{\kappa}\sin{\omega t}.
\label{eq3}
\end{eqnarray}
\noindent
Under this boundary condition, $T$ is expressed as follows: 
\begin{eqnarray}
T = T_{\rm P}+T_{\rm J},
\label{eq4}
\end{eqnarray}
\begin{eqnarray}
\left \{
\begin{array}{l}
T_{\rm P} \equiv \dfrac{ST_{0}J_{0}\sin{\omega t}}{\kappa}  \dfrac{1-i}{2\beta} \dfrac{\mathrm{e}^{\pm (1+i)\beta x}-\mathrm{e}^{\mp (1+i)\beta x}}{\mathrm{e}^{-(1+i)\beta L/2}+\mathrm{e}^{(1+i)\beta L/2}}, \\ \\
T_{\rm J}  \equiv \dfrac{\red{D} \rho J_{0}^{2}}{\kappa} \left(t - \dfrac{\sin{2\omega t} }{2\omega} \right),
\end{array}
\right.
\label{eq5}
\end{eqnarray}
where, $\beta \equiv (\omega/2\red{D})^{1/2}$ and $i$ denote the reciprocal of thermal diffusion length and an imaginary number. $T_{\rm P}$ and $T_{\rm J}$ are defined as contribution of the Peltier and Joule effects.

\noindent
A characteristic dimensionless parameter, $F\equiv(L/2)^{2}f/ 2\red{D}$, was introduced to represent the thermal response of the sample, which can classify the behavior of the temperature distribution \cite{Okawa17}. Figure~\ref{fig2}(a) shows a contour plot of the temperature change in the $x/(L/2)$-$F$ plane when the root-mean-square (rms) ac current is applied. The solid lines represent isothermal lines. The temperature changes were normalized to one of the sample edges $(x = -L/2)$. Evidently, a temperature difference $\Delta T$ in the opposite direction in the sample occurred when $F\sim1$. The direction of the $\Delta T$ was periodically inverted as the polarity of the current was reversed. Such a non-uniform $T(x,t)$ represents a thermal phase delay with respect to the current, which corresponds to insufficient cancellation of the TE effect because the phase of the thermal wave could not match the thermal response of the sample. Thus, measurements using the four-probe configuration in this region have a significant effect on the measured resistance and is among the main primary sources of error in  the $zT$ evaluation, as shown in Fig.~\ref{fig2}(b). An ac current of a sufficiently high frequency can cancel the TE effect (i.e., $F \gtrsim 15$) to yield, a flat $\Delta T$ for the sample end. Consequently, sufficiently accurate measurements can be performed by connecting the voltage terminals at this flat temperature position. 

\subsection{Exact solution of the resistance for $zT$ estimation}
Subsequently, using the obtained $T$($x$,$t$), the measured resistance $R$ can be expressed as follows, with ohmic resistance $R_{\rm ohm} \equiv \rho L_{\rm V}/A$ and resistance contributed from the Joule effect $R_{\rm J} \equiv 2ST_{\rm J}(x = L_{\rm V}/2)/I$:
\begin{eqnarray}
R = R_{\rm ohm} \left \{1+zT_{0} \left( R_1 + iR_{2} \right) + R_{\rm J} \right \},
\label{eq6}
\end{eqnarray}
\noindent
where $R_{1}$ and $R_{2}$ are expressed as the following functions:
\begin{widetext}
\begin{eqnarray}
R_{1,2} \equiv \frac{{\cos \mu \cosh \mu}(\sin \nu \cosh \nu \pm \cos \nu \sinh \nu)+\sin \mu  \sinh \mu(\sin \nu \cosh \nu \mp \cos \nu \sinh \nu)}{2\nu[(\cos \mu \cosh \mu)^2+(\sin \mu \sinh \mu)^2]},
\label{eq7}
\end{eqnarray}
\end{widetext}
\noindent
where $\mu \equiv (2\pi F)^{1/2}$ and $\nu \equiv \mu L_{\rm V}/L$ are defined as functions that are dependent on $F$ and $L_{\rm V}$, respectively. To perform the $zT$ evaluation, Eq.~(\ref{eq1}) can be rewritten as follows:
\begin{widetext}
\begin{eqnarray}
\dfrac{R_{\rm dc}-R_{\rm ac}}{R_{\rm ac}} = \dfrac{zT_{0}(1-zT_{0}R_{2})-(1-R_{\rm J})(zT_{0}R_{1}-R_{\rm J})-i\{zT_{0}R_{2}(1+zT_{0})\}
}
{(zT_{0})^{2}(R_{1}^{2}+R_{2}^2)+(1+R_{\rm J})\{2zT_{0}R_{1}+1+R_{\rm J}\}
}.
\label{eq8}
\end{eqnarray}
\end{widetext}

\noindent
Equation~(\ref{eq8}) is a general expression for $zT$ estimation using the ac Harman method with a four-probe configuration, considering the Joule effect in the time domain. It can appropriately explain the frequency characteristics of $zT$ measured with four-probe configuration. See Appendix for further calculation details.
\red{In Eq. (\ref{eq8}), $(R_{\rm dc}-R_{\rm ac})/R_{\rm ac}$  depends on the TE properties of the samples ($S, \rho, \kappa$) and the characteristic parameters ($\mu(F)$ and $\nu(F, L_{\rm V})$). If the thermal diffusivity $D$, sample length $L$, and voltage terminal distance $L_{\rm V}$ are known in advance, the characteristic parameters $F \equiv (L/2)^{2}f/2D, \mu \equiv (2\pi F)^{1/2}$, and $\nu \equiv \mu L_{\rm V}/L$ can be determined. Generally, for accurate $zT$ estimation, the current frequency $f$ should be chosen such that the dimensionless parameter $F \gtrsim 15$, as shown in Fig. 2. In our model, $zT$ can be estimated at any voltage terminal distance using Eq. (8). Furthermore, if the $f$ dependence of $R_{\rm ac}$ can be measured, Eq. (8) can be used to estimate the thermoelectric parameters ($S$ and $\kappa$) through fitting. The thermoelectric parameters ($S$ and $\kappa$) are the fitting parameters. The electrical resistivity $\rho$ is estimated using the measured $R_{\rm ac}$ and sample length $L$. }

\section{Experimental setup}
Experiments were performed using bulk samples to validate the proposed analytical model. The measurements were performed using a sintered polycrystalline sample of p-type Bi$_{0.3}$Sb$_{1.7}$Te$_{3}$ (Toshima Manufacturing Co., Ltd.)\cite{OkawaJJAP}. The Bi–Te and Sb–Te powders were sintered using the hot isostatic press (HIP) method to obtain ingots. The dimensions of the samples are $\SI{16}{mm} \times \SI{4}{mm} \times \SI{1}{mm}$. Au/Ni electrodes were fabricated via sputtering to sufficiently secure the Peltier heat generated at the edge of the sample. The heat loss from conduction through the wiring was suppressed by reducing the thickness of the Au wires, which adhered to the sample with Ag paste. To suppress the influence of the heat convection, the apparatus was assembled in a vacuum chamber, and measurements were performed at a high vacuum of $\SI{e-3}{Pa}$ or less. The sample was suspended using the thin Au wires of sufficient length to improve its thermal isolation conditions. Additionally, the sample space was covered with a radiation shield to reduce heat loss owing to thermal radiation. To realize quantitative correction of the radiation effects, measurements can be performed at higher temperatures and corrected for the measurements of other TE properties \cite{Ao18}. The ac and dc resistances of the samples were measured using an impedance analyzer. A coaxial cable was used to connect the case, and measurements were performed using a four-probe configuration. 

\section{Results and discussion}
\subsection{Current-frequency dependence of the resistance}
\begin{figure}[]
\centering
\includegraphics[width=7.5cm,clip]{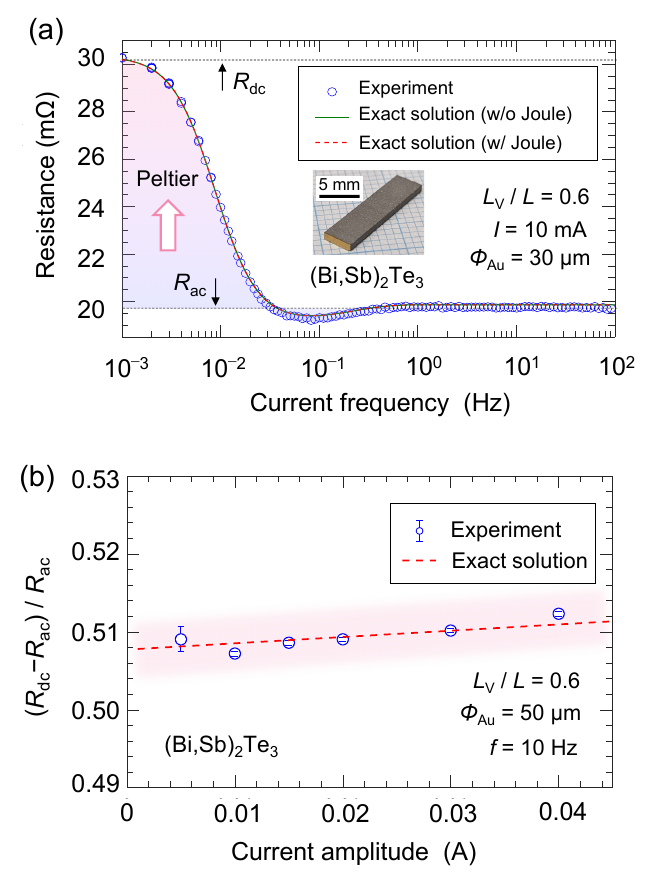}
\caption{(a) Experimental results of frequency dependence of the resistance of (Bi,Sb)$_{2}$Te$_{3}$. Open symbols correspond to experimental results with $I = \SI{10}{mA}$, $\phi_{\rm Au} = \SI{30}{\micro m}$, and $L_{\rm V}/L$ = 0.6. Red and green lines represent calculation results based on the exact solution with and without Joule effect, respectively. Inset displays a photograph of the sample. (b) Relationship between the $(R_{\rm dc}-R_{\rm ac})/R_{\rm ac}$ and current amplitude. Dotted lines show the exact solution with the Joule effect. Open symbols are the experimental results at $f = \SI{10}{Hz}$, $\phi_{\rm Au} = \SI{50}{\micro m}$, and $L_{\rm V}/L$ = 0.6. The error bars show the standard deviation of the mean for repeated measurements of the resistance.}
\label{fig3}
\end{figure}

We demonstrate the correctness of the exact solution and $zT$ evaluation using a four-probe configuration. Figure~\ref{fig3}(a) shows the experimental and calculated results of the current-frequency dependence of the resistance for (Bi,Sb)$_{2}$Te$_{3}$. The open symbols represent the experimental results for $I = \SI{10}{mA}$ and a wire diameter of $\phi_{\rm Au}= \SI{30}{{\micro} m}$. The normalized distance between the voltage terminals $L_{\rm V}/L$ is 0.6. The calculation used the typical physical properties of (Bi,Sb)$_{2}$Te$_{3}$ at room temperature $(\red{D} = \SI{1.25e-6}{m^2/s}, S = \SI{166}{\micro V/K}, \rho = \SI{1.1e-5}{\Omega m}, \kappa = \SI{1.45}{W/mK})$. At current frequencies less than $\SI{1}{mHz}$, the measured resistance approaches $R_{\rm dc}$, and the TE effect was sufficiently generated. However, it gradually decreased to $R_{\rm ac}$ as the current frequency $f$ increased. Moreover, a dip structure occurs at $f \sim \SI{80}{mHz}$, corresponding to $F \sim 1$ in Fig.~\ref{fig2}. It is caused by a phase shift owing to the thermal wave and thermal response of the sample. The voltage is detected at the points $(x = \pm L_{\rm V}/2)$ where the temperature change is opposite direction to the temperature difference between the two ends of the sample $(x = \pm L/2)$. This is referred to as the “thermoinductive effect”\cite{Okawa17}. As described later, this resulted in the overestimation of $zT$, which is evident from the $F$ dependence of $(R_{\rm dc}-R_{\rm ac})/R_{\rm ac}$. Such a dip structure is not observed in the two-probe configuration when $L_{\rm V}/L = 1$; however, it is underestimated compared to $zT$ unless a sufficiently high frequency is chosen (Fig.~\ref{fig2}(b)). This is consistent with the lack of a dip structure in two-probe measurements \cite{Shinozaki19,Hasegawa20,Hasegawa21,Marchi22,Marchi23,Apertet24,Pitarch25,Pitarch26,Arisaka27,Pitarch28,Yoo29,Otsuka30,Zaoui31,Hasegawa32,Pitarch33,Hirabayashi34,Korzhuev35}. Thus, the ac Harman method can be used to determine $zT$ over a wide frequency range of the four-probe configuration using the exact solution derived. 
\subsection{Influence of the Joule effect}
Subsequently, the influence of the Joule effect was investigated. The red and green lines shown in Fig.~\ref{fig3}(a) represent the calculation results based on the exact solution with and without considering the Joule effect, respectively. As Joule heating is time-dependent, the measured resistance was calculated over the time of one measurement cycle at each frequency. When $I = \SI{10}{mA}$, the influence of the Joule effect was not observed at most frequencies. Figure~\ref{fig3}(b) illustrates the relationship between $(R_{\rm dc}-R_{\rm ac})/R_{\rm ac}$ and the current amplitude. The experimental results are shown for current values in the range of $\SI{5}{mA}$ to $\SI{40}{mA}$ \red{with $f = \SI{10}{Hz}$, $\phi_{\rm Au} = \SI{50}{\micro m}$, and $L_{\rm V}/L$ = 0.6}. The error bars represent the standard deviation of the mean obtained from independent measurements. $(R_{\rm dc}-R_{\rm ac})/R_{\rm ac}$ increases linearly with the current amplitude. The result is consistent with the calculation line based on the exact solution. 

\begin{figure}[]
\centering
\includegraphics[width=8cm,clip]{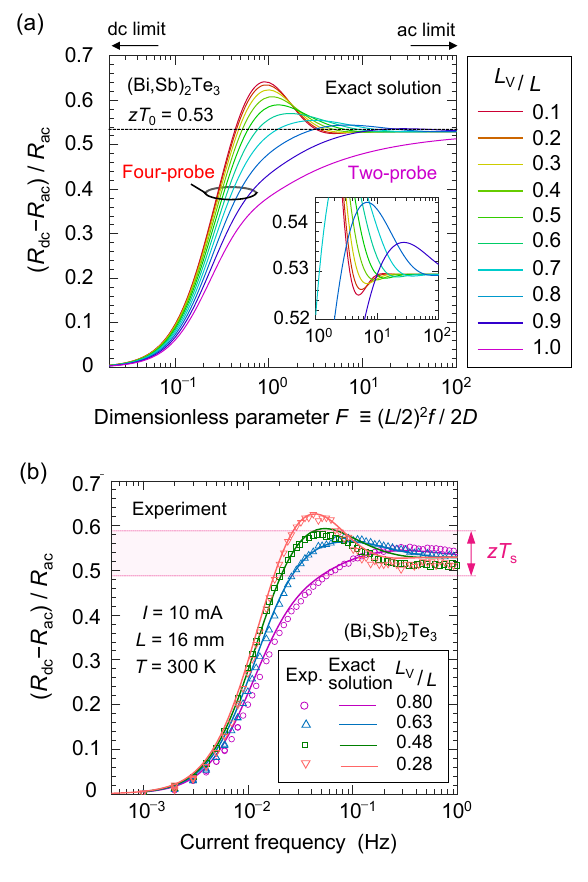}
\caption{(a) Calculation results based on the exact solution of $F$ dependence of $(R_{\rm dc}-R_{\rm ac})/R_{\rm ac}$ for the variation of the $L_{\rm V}/L$ of (Bi,Sb)$_{2}$Te$_{3}$. (b) Experimental results of the frequency dependence of $(R_{\rm dc}-R_{\rm ac})/R_{\rm ac}$. Open symbols represent experimental results, and solid lines represent calculation results. Purple circles, blue triangles, green squares, and orange reverse triangles correspond to the experimental results for $L_{\rm V}/L$ = 0.80, 0.63, 0.48, and 0.1, respectively. Solid lines represent the calculation results of the exact solution. $zT$ estimated using standard formula $zT_{\rm s} \equiv S^2\red{T}/(\rho \kappa) = 0.538 \pm \SI{10}{\%}$ is  shown as the red area.}
\label{fig4}
\end{figure}

At $I = \SI{10}{mA}$, $(R_{\rm dc}-R_{\rm ac})/R_{\rm ac}$ was estimated to be 0.509, which increased to 0.511 at $I = \SI{40}{mA}$. If the higher order TE effect, the Thomson heat induced by the Joule effect, $(R_{\rm dc}-R_{\rm ac})/R_{\rm ac}$ would be proportional to the square of the current at high current region. Upon applying a large current of $\SI{50}{mA}$ or more, the sample and wiring heat up, which renders accurate evaluation and correction difficult. The optimal current amplitude can be determined by comparing the Peltier heat generated in the measurement system with the Joule heat and then choosing a suitable compromise between the degree of measurement noise generated by the setup and Joule heat \cite{Hasegawa32}.

\begin{figure}[]
\centering
\includegraphics[width=7.5cm,clip]{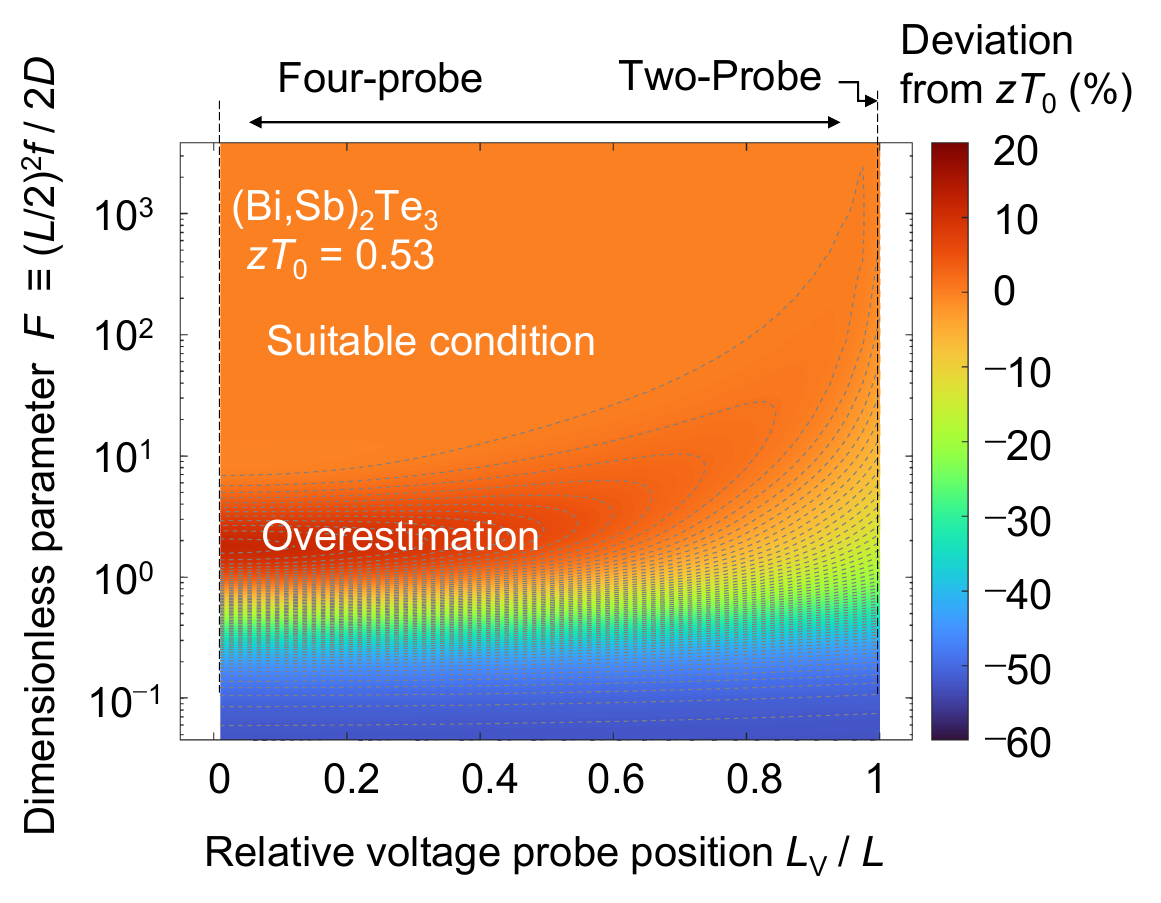}
\caption{Contour plot of the deviation from $zT_{0}$ in $F$-$L_{\rm V}/L$ plane. $zT_{0}$ of (Bi,Sb)$_{2}$Te$_{3}$ is estimated using the typical physical parameters at room temperature as describe in the text.}
\label{fig5}
\end{figure}

\subsection{Suitable measurement conditions for the ac Harman method with four-probe configuration}
Figure~\ref{fig4} presents the experimental and calculated results based on the exact solution of $(R_{\rm dc}-R_{\rm ac})/R_{\rm ac}$ for different $L_{\rm V}/L$ values. Figure \ref{fig4}(a) shows the dependence of $F$ on the exact solution of $(R_{\rm dc}-R_{\rm ac})/R_{\rm ac}$, where a smaller and larger $F$ could be considered as dc and ac limits, respectively. $L_{\rm V}/L$ was varied from 0.1 to 1.0. For an $F$ value of approximately 1, the smaller the $L_{\rm V}/L$, the larger the dip structure. It can be attributed to the thermal phase delay (as a thermoinductive effect) and is a measurement error that only occurs in the four-probe configuration, as described above. The intrinsic value of dimensionless figure of merit $zT_{0} = 0.53$ can be determined using $R_{\rm ac}$ at the ac limit. The value of $(R_{\rm dc}-R_{\rm ac})/R_{\rm ac}$ changes significantly with respect to $L_{\rm V}/L$, and it is evident that the maximum deviation is $\SI{20}{\%}$. It was quantitatively shown that the choice of the current frequency causes an error in $zT$ evaluation for the ac Harman method.

Figure~\ref{fig4}(b) illustrates the experimental results of the frequency dependence of $(R_{\rm dc}-R_{\rm ac})/R_{\rm ac}$. $L_{\rm V}/L$ was varied from 0.28 to 0.80. For the $L_{\rm V}/L$ value of 0.28, a large dip structure was observed at approximately $\SI{40}{mHz}$. For all $L_{\rm V}/L$ results, consistency was obtained with the exact solution over a wide frequency range. As the dimensionless resistance $R_{1,2}$ in Eq.~(\ref{eq8}) depends on $F$, the current frequency to be selected differs based on sample length $L$ and thermal diffusivity $\red{D}$. Therefore, it is not always necessary to select an appropriate current frequency when measuring samples of different sizes and compositions. However, for an ac bridge used for accurate resistance measurement, the selection of the measurement frequency must be focused upon. When measuring higher frequencies, the influence of the parasitic effects, inherent to the high frequency, depends on the circuit; therefore, the frequency must be checked for each instance. As depicted in Fig.~\ref{fig5}, the suitable experimental condition for the ac Harman method was identified visually.

To verify the estimated $zT$ value, independent measurements of the Seebeck coefficient, electrical resistivity, and thermal conductivity were performed at 300 K. The rectangular bars cut from the ingots were used to measure the electrical resistivity ($\rho$) and the Seebeck coefficient ($S$) using ZEM-3 (ULVAC Co., Ltd.). The power factor, defined as $S^{2}/\rho$, was calculated ($S=\SI{158}{\micro V/K}, \rho = \SI{6.6e-6}{\Omega m}$). The materials demonstrate a relatively high power-factor of approximately $\SI{3.8}{mW {mK}^{-2}}$ at $\SI{300}{K}$. The thermal conductivity ($\kappa$) of the materials was determined using the formula  $\kappa = \red{D} d C_{\rm p}$, in which $\red{D}$ is the thermal diffusivity, $d$ represents the density, and $C_{\rm p}$ denotes the specific heat of the materials. The thermal diffusivity was measured via a laser flash method $(\red{D} = \SI{1.6e-6}{\red{m^{2}}/s})$, and the density was measured using the Archimedes method $(d = \SI{6.64} {g/cm^{3}})$. The specific heat of the material was determined to be $C_{\rm p} = \SI{0.199}{J/gK}$ based on the literature values for Bi$_{2}$Te$_{3}$ ($\SI{124.4}{J/Kmol}$) and Sb$_{2}$Te$_{3}$ ($\SI{128.8}{J/Kmol}$) \cite{ChenCp}. The thermal conductivity was calculated using the measured $\red{D}$, $d$, and the estimated $C_{\rm p} (\kappa = \SI{2.11} {W/mK})$. The dimensionless figure of merit was calculated using the standard formula $zT_{\rm s} = S^{2}\red{T}/(\rho \kappa)$ and is 0.538. In a report on the international round-robin test using bulk bismuth telluride, the scatter for $zT$ was estimated to be approximately $\SI{10}{\%}$ at 300 K \cite{Wang5}. The typical error in estimating $zT$ from resistance measurement using the Harman formula $(R_{\rm dc}-R_{\rm ac})/R_{\rm ac}$ is reported to be about $\SI{15}{\%}$ \cite{BoorVDP}. The $zT$ evaluation results using the standard formula are in agreement with those obtained using the ac Harman method, as shown in Fig. 4(b).

\begin{figure}[]
\centering
\includegraphics[width=7cm,clip]{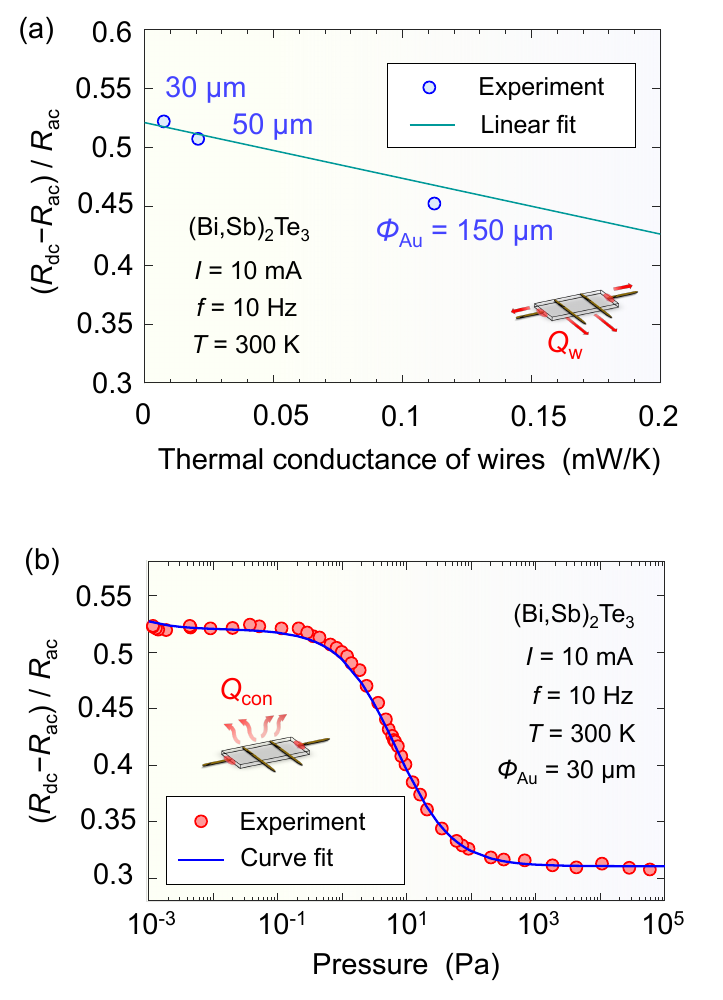}
\caption{(a) Results of the thermal conductance of the lead wires as a function of $(R_{\rm dc}-R_{\rm ac})/R_{\rm ac}$ with $I =\SI{10}{mA}, f=\SI{10}{Hz}$, and $T = \SI{300}{K}$. The blue circles denote the experimental results when the Au wire diameter is changed from $\phi_{\rm Au} = \SI{30}{\red{\micro} m}$ to $\SI{150}{\red{\micro} m}$. The green line denotes the linear fit line. (b) Pressure dependence results of $(R_{\rm dc}-R_{\rm ac})/R_{\rm ac}$ using (Bi,Sb)$_{2}$Te$_{3}$ with $I =\SI{10}{mA}, f=\SI{10}{Hz}$, $\phi_{\rm Au} = \SI{30}{\micro m}$, and $T = \SI{300}{K}$. The red circles correspond to the experimental results. The blue line corresponds to the calculation results.}
\label{fig6}
\end{figure}

\subsection{Examination of accuracy and influence of other error factors}
Thus far, we presented error factors such as the current frequency $f$, current amplitude $I$, and voltage terminal distance $L_{\rm V}$, which have not been quantitatively identified using exact solutions. Subsequently, we discuss the other significant error factors, excluding dimensionless parameter $F$, $L_{\rm V}$, and the Joule effect. The ac Harman method assumes that all Peltier heat generated at the sample edge flows into the sample. In an actual experimental environment, various heat losses occur, such as conduction through the lead wire $Q_{\rm w}$, and heat convection $Q_{\rm con}$, which results in a crucial error factor in the $zT$ estimation \cite{Kwon37,Penn38,Campbell39,Iwasaki40,Kang41,Roh42,Putilin43, Maccarty44,Vasilevskiy45}. The exact solutions comprehensively consider $Q_{\rm w}$ and $Q_{\rm con}$ combined with other TE effects; however, it is extremely difficult to introduce using the unsteady heat conduction equation. Our results of the resistance measurements for different wire diameters $\phi_{\rm Au}$ with $I =\SI{10}{mA}, f = \SI{10}{Hz}$, and $T = \SI{300}{K}$ are shown in Fig.~\ref{fig6}(a). Considering an energy balance\cite{Kwon37}, the ratio of the thermal conductance of the lead wire $K_{\rm w}$ to that of the sample $K$ determines $Q_{\rm w}$. The results of the resistance measurements with wire diameter $\phi_{\rm Au}$ in the range of $\SI{30}{\micro m}$ to $\SI{150}{\micro m}$ exhibits that $(R_{\rm dc}-R_{\rm ac})/R_{\rm ac}$ is underestimated as $K_{\rm w}$ increases. The result is quantitatively consistent with the relationship expressed as $(R_{\rm dc}-R_{\rm ac})/R_{\rm ac} \propto 1/K_{\rm w}$ derived by a linear fitting. Thus, reduction in $\phi_{\rm Au}$ can alleviate the underestimation of $zT$ owing to the $Q_{\rm w}$. A wire with a smaller diameter has the effect of lowering the upper limit of the maximum applied current, which is a trade-off in an actual experimental environment.
\begin{figure}[]
\centering
\includegraphics[width=7.5cm,clip]{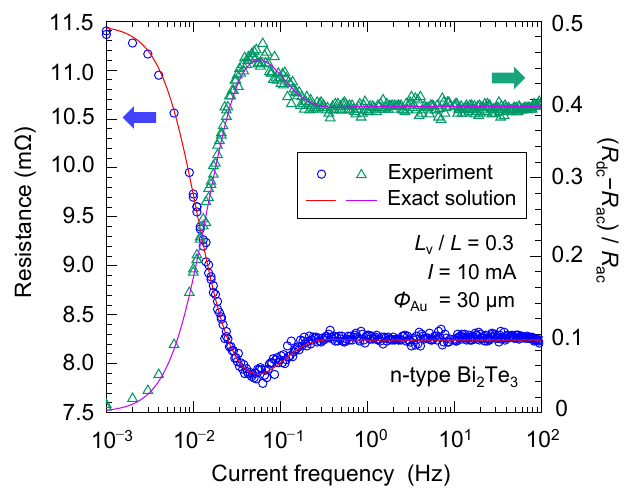}
\caption{\red{Experimental results of frequency dependence of the resistance and $(R_{\rm dc}-R_{\rm ac})/R_{\rm ac}$ of n-type Bi$_{2}$Te$_{3}$. Open symbols correspond to experimental results with $L_{\rm V}/L$ = 0.3, $I = \SI{10}{mA}$ and $\phi_{\rm Au} = \SI{30}{\micro m}$. Solid lines represent calculation results based on the exact solution.}}
\end{figure}
As shown in Fig.~\ref{fig6}(b), resistance measurements were performed by varying the pressure in the sample chamber from ambient pressure to $\SI{e-3}{Pa}$ with $I =\SI{10}{mA}, f=\SI{10}{Hz}$, $\phi_{\rm Au} = \SI{30}{\red{\micro} m}$, and $T = \SI{300}{K}$. At ambient pressure, $(R_{\rm dc}-R_{\rm ac})/R_{\rm ac}$ can be underestimated by up to $\SI{40}{\%}$ compared with that in a high vacuum. A sharp decrease in $(R_{\rm dc}-R_{\rm ac})/R_{\rm ac}$ was observed from $\SI{1}{Pa}$ to $\SI{10}{Pa}$. This can be attributed to the transition from the viscous to the molecular flow regime, resulting in an increase in the number of molecules responsible for heat conduction, leading to increased convective heat loss $Q_{\rm con}$. Analytically, the introduction of the heat transfer term $hP(T-T_{0})$ within the heat conduction equation yields the relationship $(R_{\rm dc}-R_{\rm ac})/R_{\rm ac} \propto  \tanh{(hPL/2\kappa A)}^{1/2}$, where $h$ and $P$ are the heat-transfer coefficient and perimeter of the sample, respectively. Fitting with this relationship yielded results that are consistent with the experimental results over a wide range of pressures. For our setup, the influence of $Q_{\rm con}$ can be neglected in a vacuum environment above $\SI{e-1}{Pa}$.

The existence of  inhomogeneous Peltier heating at the edge of the sample/electrode junctions also causes a $\SI{12}{\%}$ underestimation for $zT$ \cite{Pitarch36}. In our setup, the effect of inhomogeneous current application was avoided through the formation of Ni/Au electrodes at both ends of the (Bi,Sb)$_{2}$Te$_{3}$. Furthermore, the heat radiation leads to significant underestimation up to \SI{9}{\%} at high temperatures, $\SI{440}{K}$ \cite{Ao18}. With proper installation of radiation shielding, the influence of the radiation heat loss can be reduced at room temperature.

The primary error factors were identified as described above. The insights gained in this analysis enable accurate $zT$ estimation measured using four-probe configuration measurements. Notably, some known physical parameters are required to fully compensate for error factors, which is also true for other evaluation methods. For example, in the two-probe based impedance measurement, which is an accurate measurement technique with a similar concept, the frequency dependence of impedance is measured in advance over a wide range \cite{Shinozaki19,Hasegawa20,Hasegawa21,Marchi22,Marchi23,Apertet24,Pitarch25,Pitarch26,Arisaka27,Pitarch28,Yoo29,Otsuka30,Zaoui31,Hasegawa32,Pitarch33,Hirabayashi34,Korzhuev35}. Consequently, several thermophysical properties can be estimated using analytical models in the frequency domain. Since the TE module system tends to be relatively complex within the time domain, the analysis is used within the frequency domain \cite{Zaoui31}. In the two-probe configuration, the contact resistance between the sample and electrode must be considered, which necessitates its analysis. Furthermore, the influence of contact resistance is inevitable. Experimental studies, such as preparing a sample shape, wherein this influence can be ignored, are necessary \cite{Pitarch36}. For accurate resistance measurement, the use of a four-probe configuration that  requires knowledge of the voltage terminal distance is beneficial. Note that the error in the evaluation of the voltage terminal distance and the thermal diffusivity has an influence on our analysis. The measurement of the voltage terminal distance is more influenced by shorter sample lengths or larger wire diameters. It has been suggested that the error of length measurement tend to be quite large ($\geqq \SI{10}{\%}$) in small samples \cite{HeremansNM}.

\subsection{$zT$ estimation of other sample}
To demonstrate the universality of our technique, we have also applied the ac Harman method to estimate the $zT$ of n-type Bi$_2$Te$_3$ (Toshima Manufacturing Co., Ltd.) using the same sample geometry as the p-type Bi$_{0.3}$Sb$_{1.7}$Te$_{3}$. Figure 7 shows the frequency dependence of the resistance and $(R_{\rm dc}-R_{\rm ac})/R_{\rm ac}$ of n-type Bi$_{2}$Te$_{3}$ with $I = \SI{10}{mA}$ and $\phi_{\rm Au} = \SI{30}{\micro m}$. The normalized distance between the voltage terminals $L_{\rm V}/L$ is 0.3. Notable dip structures were observed around 50 mHz in the frequency dependences of the resistance and $(R_{\rm dc}-R_{\rm ac})/R_{\rm ac}$. This frequency region roughly corresponds to $F = 1$. The plots developed using the exact solution, agreed well with the experimental results over a wide frequency range.  The value of dimensionless figure of merit $zT_{0} = 0.41$ can be determined using the measurement data. The physical properties of n-type Bi$_2$Te$_3$ ($D = \SI{1.6e-6}{m^{2}/s}$, $S = \SI{-133}{\micro V/K}$, $\rho = \SI{6.2e-6}{\Omega m}$ and $\kappa = \SI{2.11}{W/mK}$) at $\SI{300}{K}$ were used in the calculation. As with the case of p-type Bi$_{0.3}$Sb$_{1.7}$Te$_{3}$, the Seebeck coefficient and electrical resistivity were measured using ZEM-3 and found to be $\SI{-131}{\micro V/K}$ and $\SI{5.9e-6}{\Omega m}$, respectively. The dimensionless figure of merit using the standard formula $zT_{\rm s}$ is calculated 0.39, in which the thermal conductivity is assumed the same value as the p-type Bi$_{0.3}$Sb$_{1.7}$Te$_{3}$ ($\kappa = \SI{2.11}{W /mK}$). The $zT$ evaluation results using the standard formula are in agreement with those obtained using the ac Harman method. The present analysis is applicable to other thermoelectric materials.

\section{Conclusion}
In this study, the evaluation equation of the ac Harman method was derived for a four-probe configuration from the exact solution, which was then experimentally verified. The effects of the distance between the voltage terminals, current frequency, Joule effect, and convective heat transfer, which have been empirically avoided, were clarified. Using the evaluation formula, each error factor was avoided or corrected. The analysis performed provided information on $zT$ evaluation errors caused by the position of the voltage terminal attachments, as well as heat losses owing to thermal convection, heat conduction of the lead wires, and the Joule effect.  As described above, this study realized the correction of various heat losses that are inevitable when measuring $zT$ using the ac Harman method, and a more accurate determination of $zT$ can be realized through further systematic empirical research.

\section*{Acknowledgements}
This work was supported by a Grant-in-Aid for Research Activity Start-up (Grant No.17H07399) from the Japan Society for Promotion of Science (JSPS), Grant-in-Aid for Early-Career Scientists (Grant No.23K13552) from JSPS, the Thermal \& Electric Energy Technology Foundation (TEET), Iketani Science and Technology Foundation, and the Precise Measurement Technology Promotion Foundation (PMTP-F).

\begin{widetext}
\section*{Appendix: Calculation details for exact solution of $zT$ estimation}
The temperature distribution in the sample $T(x,t)$ at position $x$ and time $t$ is expressed as Eq.~(4) and Eq.~(5). The contribution from the Peltier effects $T_{\rm P}(x,t)$ can be obtained by solving the following one-dimensional unsteady-state heat transfer equation \cite{Okawa17,Kirby}:
\begin{eqnarray}
\frac{\partial^2 T_{\rm P}(x,t)}{\partial x^2}=\frac{1}{\red{D}}\frac{\partial T_{\rm P}(x,t)}{\partial t}.
\end{eqnarray}
According to the method of separation of variables, the general solution of Eq.~(9) is described as
\begin{eqnarray}
T_{\rm P}(x,t) = B_{1} \mathrm{e}^{\pm(1+i)\beta x-i\omega t}+B_{2}\mathrm{e}^{\mp (1+i)\beta x-i \omega t}, 
\end{eqnarray}
where $B_{1}$ and $B_{2}$ are arbitrary constants. $\beta \equiv (\omega/2\red{D})^{1/2}$ is the reciprocal of thermal diffusion length $D_{\rm th} = (\red{D}/\pi f)^{1/2}$. Under the boundary condition Eq. (3), $B_{1}$ and $B_{2}$ are obtained.
\begin{equation}
\begin{pmatrix}
\displaystyle (1+i)\beta \mathrm{e}^{(1+i)\beta L/2-i\omega t} & -(1+i)\beta \mathrm{e}^{-(1+i)\beta L/2 -i\omega t} \\
\\
(1+i)\beta \mathrm{e}^{-(1+i)\beta L/2-i\omega t} & -(1+i)\beta \mathrm{e}^{(1+i)\beta L/2 -i\omega t}
\end{pmatrix} 
\begin{pmatrix}
B_{1} \\
\\
B_{2}
\end{pmatrix} 
=
\begin{pmatrix}
 \dfrac{ST_{0}J_{0}}{\kappa}\sin{\omega t} \\
\\
 \dfrac{ST_{0}J_{0}}{\kappa}\sin{\omega t}
\end{pmatrix} ,
\end{equation}
\begin{equation}
\left \{
\begin{array}{l}
B_{1} = \dfrac{ST_{0}J_{0}\sin{\omega t}}{\kappa}\dfrac{1+i}{2i\beta}\dfrac{\mathrm{e}^{i\omega t}}{\mathrm{e}^{-\beta L/2(1+i)}+\mathrm{e}^{\beta L/2(1+i)}}\\
\\
B_{2} = \dfrac{ST_{0}J_{0}\sin{\omega t}}{\kappa}\dfrac{1+i}{2i\beta}\dfrac{-\mathrm{e}^{i\omega t}}{\mathrm{e}^{-\beta L/2 (1+i)}+\mathrm{e}^{\beta L/2(1+i)}}
\end{array}
\right. .
\end{equation}
$T_{\rm P}(x,t)$ is given by Eq. (5). The contribution from the Joule effects $T_{\rm J}(x,t)$ can be obtained by solving the following equation under the boundary condition \cite{DunnThomson, Dames3omega, Lu3omega}:
\begin{eqnarray}
\frac{\partial^2 T_{\rm J}(x,t)}{\partial x^2}=\frac{1}{\red{D}}\frac{\partial T_{\rm J}(x,t)}{\partial t}-\dfrac{\rho (J_{0}\sin{\omega t})^{2}}{\kappa},   \qquad
\left.\frac{\partial T_{\rm J}(x,t)}{\partial x}\right|_{x=\pm L/2} = 0.
\end{eqnarray}
Using $T = T_{\rm P} +T_{\rm J}$ obtained from Eq. (5), the voltage $V$ measured between $\pm L_{\rm V}/2$ is expressed as follows:
\begin{eqnarray}
V &=& L_{\rm V} \rho J_{0}\sin{\omega t} + \int_{-L/2}^{L/2} S\dfrac{\partial T(x,t)}{\partial x}dx \nonumber \\ 
&=& L_{\rm V} \rho J_{0}\sin{\omega t} + 2ST(L_{\rm V}/2,t) \nonumber \\  
&=& L_{\rm V} \rho J_{0}\sin{\omega t}  + 2S\left \{ \frac{ST_{0}J_{0}\sin{\omega t}}{\kappa}\dfrac{1-i}{2\beta} \dfrac{\mathrm{e}^{\pm (1+i)\beta L_{\rm V}/2}-\mathrm{e}^{\mp (1+i)\beta L_{\rm V}/2}}{\mathrm{e}^{-(1+i)\beta L/2}+\mathrm{e}^{(1+i)\beta L/2}} + \dfrac{\red{D} \rho J_{0}^{2}}{\kappa} \left(t - \dfrac{\sin{2\omega t} }{2\omega} \right) \right \}. 
\end{eqnarray}
The first term on the right side represents the ohmic voltage between $L_{\rm V}$, and the second term represents the Seebeck voltage. In actual measurements, the
temperature rise caused by the Peltier and Joule effects at low current region is considered small $(\sim \SI{0.1}{K})$, so the
integral part can be replaced with $2ST(x = L_{\rm V}/2,t)$. The measured resistance $R$ in Eq. (6) can be introduced as $R = V/I$:
\begin{eqnarray}
R &=& \dfrac{V}{I} = \dfrac{V}{JA}  \nonumber \\ 
&=& \rho \dfrac{L_{\rm V}}{A}  + 2S\left \{ \frac{ST_{0}}{\kappa A}\dfrac{1-i}{2\beta} \dfrac{\mathrm{e}^{\pm (1+i)\beta L_{\rm V}/2}-\mathrm{e}^{\mp (1+i)\beta L_{\rm V}/2}}{\mathrm{e}^{-(1+i)\beta L/2}+\mathrm{e}^{(1+i)\beta L/2}} + \dfrac{\red{D} \rho J_{0}}{\kappa A\sin{\omega t}} \left(t - \dfrac{\sin{2\omega t} }{2\omega} \right) \right \} \nonumber \\
&=& R_{\rm ohm}\left\{1+  \frac{S^{2}T_{0}}{\kappa \rho}\dfrac{1-i}{2\beta L_{\rm V}/2} \dfrac{\mathrm{e}^{\pm (1+i)\beta L_{\rm V}/2}-\mathrm{e}^{\mp (1+i)\beta L_{\rm V}/2}}{\mathrm{e}^{-(1+i)\beta L/2}+\mathrm{e}^{(1+i)\beta L/2}} + \dfrac{2S\red{D} J_{0}}{\kappa L_{\rm V}\sin{\omega t}} \left(t - \dfrac{\sin{2\omega t} }{2\omega} \right) \right \} .
\end{eqnarray}
The equation can be rewritten using $\mu \equiv \beta L/2$ and $\nu \equiv \beta L_{\rm V}/2$:
\begin{eqnarray}
R = R_{\rm ohm}\left\{1+ zT_{0} \dfrac{1-i}{2\nu} \dfrac{\mathrm{e}^{\pm (1+i)\nu}-\mathrm{e}^{\mp (1+i)\nu}}{\mathrm{e}^{-(1+i)\mu}+\mathrm{e}^{(1+i)\mu}} + \dfrac{2S\red{D} J_{0}}{\kappa L_{\rm V}\sin{\omega t}} \left(t - \dfrac{\sin{2\omega t} }{2\omega} \right) \right \},
\end{eqnarray}
where
\begin{flalign}
\dfrac{1-i}{2\nu} \dfrac{\mathrm{e}^{\pm (1+i)\nu}-\mathrm{e}^{\mp (1+i)\nu}}{\mathrm{e}^{-(1+i)\mu}+\mathrm{e}^{(1+i)\mu}} 
=& \dfrac{1+i}{2\nu}\dfrac{\sin{\nu}\cosh{\nu}-i\cos{\nu}\sinh{\nu}}{\cos{\mu}\cosh{\mu}+i\sin{\mu}\sinh{\mu}} \\
=& \dfrac{\cos{\mu}\cosh{\mu}(\sin{\nu}\cosh{\nu}+\cos{\nu}\sinh{\nu})+\sin{\mu}\sinh{\mu}(\sin{\nu}\cosh{\nu}-\cos{\nu}\sinh{\nu})}{2\nu(\cos{\mu}\cosh{\mu}+i\sin{\mu}\sinh{\mu})(\cos{\mu}\cosh{\mu}-i\sin{\mu}\sinh{\mu})} \nonumber \\
\hspace*{9em}& + \dfrac{i \left \{ \cos{\mu}\cosh{\mu}(\sin{\nu}\cosh{\nu}-\cos{\nu}\sinh{\nu})-\sin{\mu}\sinh{\mu}(\sin{\nu}\cosh{\nu}+\cos{\nu}\sinh{\nu} \right \} }{2\nu(\cos{\mu}\cosh{\mu}+i\sin{\mu}\sinh{\mu}(\cos{\mu}\cosh{\mu}-i\sin{\mu}\sinh{\mu})}.
\end{flalign}
Using Eq. (7), the resistance $R$ can be expressed as Eq. (6). 
In the dc limit, the current frequency $f \to 0$, $T_{\rm P}$ and $T_{\rm J}$ is expressed as:
\begin{eqnarray}
T_{\rm P} = \dfrac{ST_{0}J_{0}}{\kappa}, \quad T_{\rm J}  = \dfrac{\red{D} \rho J_{0}^{2}}{\kappa}.
\end{eqnarray}
The resistances of the sample measured via application of dc and ac currents, $R_{\rm dc}$ and $R_{\rm ac}$ are obtained as:
\begin{eqnarray}
R_{\rm dc} &=& R_{\rm ohm}\left(1+ zT_{0} + \dfrac{2S\red{D} J_{0}}{\kappa L_{\rm V}}t \right ), \\
R_{\rm ac} &=& R_{\rm ohm}\left\{1+ zT_{0} \dfrac{1-i}{2\nu} \dfrac{\mathrm{e}^{\pm (1+i)\nu}-\mathrm{e}^{\mp (1+i)\nu}}{\mathrm{e}^{-(1+i)\mu}+\mathrm{e}^{(1+i)\mu}} + \dfrac{2S\red{D} J_{0}}{\kappa L_{\rm V}\sin{\omega t}} \left(t - \dfrac{\sin{2\omega t} }{2\omega} \right) \right \}.
\end{eqnarray}
To perform the $zT$ evaluation, Eq. (1) can be rewritten as follows:
\begin{eqnarray}
\dfrac{R_{\rm dc}-R_{\rm ac}}{R_{\rm ac}} &=& \dfrac{R_{\rm dc}}{R_{\rm ac}}-1\nonumber \\
&=& \dfrac{
R_{\rm ohm}\left(1+ zT_{0} + \dfrac{2S\red{D} J_{0}}{\kappa L_{\rm V}}t \right )
}
{
R_{\rm ohm}\left\{1+ zT_{0} \dfrac{1-i}{2\nu} \dfrac{\mathrm{e}^{\pm (1+i)\nu}-\mathrm{e}^{\mp (1+i)\nu}}{\mathrm{e}^{-(1+i)\mu}+\mathrm{e}^{(1+i)\mu}} + \dfrac{2S \red{D} J_{0}}{\kappa L_{\rm V}\sin{\omega t}} \left(t - \dfrac{\sin{2\omega t} }{2\omega} \right) \right \}
}-1.
\end{eqnarray}
\end{widetext}

\nocite{*}					
\bibliography{Harman}

\begin{thebibliography}{57}%
\makeatletter
\providecommand \@ifxundefined [1]{%
 \@ifx{#1\undefined}
}%
\providecommand \@ifnum [1]{%
 \ifnum #1\expandafter \@firstoftwo
 \else \expandafter \@secondoftwo
 \fi
}%
\providecommand \@ifx [1]{%
 \ifx #1\expandafter \@firstoftwo
 \else \expandafter \@secondoftwo
 \fi
}%
\providecommand \natexlab [1]{#1}%
\providecommand \enquote  [1]{``#1''}%
\providecommand \bibnamefont  [1]{#1}%
\providecommand \bibfnamefont [1]{#1}%
\providecommand \citenamefont [1]{#1}%
\providecommand \href@noop [0]{\@secondoftwo}%
\providecommand \href [0]{\begingroup \@sanitize@url \@href}%
\providecommand \@href[1]{\@@startlink{#1}\@@href}%
\providecommand \@@href[1]{\endgroup#1\@@endlink}%
\providecommand \@sanitize@url [0]{\catcode `\\12\catcode `\$12\catcode
  `\&12\catcode `\#12\catcode `\^12\catcode `\_12\catcode `\%12\relax}%
\providecommand \@@startlink[1]{}%
\providecommand \@@endlink[0]{}%
\providecommand \url  [0]{\begingroup\@sanitize@url \@url }%
\providecommand \@url [1]{\endgroup\@href {#1}{\urlprefix }}%
\providecommand \urlprefix  [0]{URL }%
\providecommand \Eprint [0]{\href }%
\providecommand \doibase [0]{http://dx.doi.org/}%
\providecommand \selectlanguage [0]{\@gobble}%
\providecommand \bibinfo  [0]{\@secondoftwo}%
\providecommand \bibfield  [0]{\@secondoftwo}%
\providecommand \translation [1]{[#1]}%
\providecommand \BibitemOpen [0]{}%
\providecommand \bibitemStop [0]{}%
\providecommand \bibitemNoStop [0]{.\EOS\space}%
\providecommand \EOS [0]{\spacefactor3000\relax}%
\providecommand \BibitemShut  [1]{\csname bibitem#1\endcsname}%
\let\auto@bib@innerbib\@empty
\bibitem [{\citenamefont {Bell}(2008)}]{Bell1}%
  \BibitemOpen
  \bibfield  {author} {\bibinfo {author} {\bibfnamefont {L.~E.}\ \bibnamefont
  {Bell}},\ }\href@noop {} {\bibfield  {journal} {\bibinfo  {journal}
  {Science}\ }\textbf {\bibinfo {volume} {321}},\ \bibinfo {pages} {1457}
  (\bibinfo {year} {2008})}\BibitemShut {NoStop}%
\bibitem [{\citenamefont {Zhang}\ and\ \citenamefont {Zhao}(2015)}]{Zhang2}%
  \BibitemOpen
  \bibfield  {author} {\bibinfo {author} {\bibfnamefont {X.}~\bibnamefont
  {Zhang}}\ and\ \bibinfo {author} {\bibfnamefont {L.-D.}\ \bibnamefont
  {Zhao}},\ }\href@noop {} {\bibfield  {journal} {\bibinfo  {journal} {J.
  Materiomics}\ }\textbf {\bibinfo {volume} {1}},\ \bibinfo {pages} {92}
  (\bibinfo {year} {2015})}\BibitemShut {NoStop}%
\bibitem [{\citenamefont {He}\ and\ \citenamefont {Tritt}(2017)}]{He3}%
  \BibitemOpen
  \bibfield  {author} {\bibinfo {author} {\bibfnamefont {J.}~\bibnamefont
  {He}}\ and\ \bibinfo {author} {\bibfnamefont {T.~M.}\ \bibnamefont {Tritt}},\
  }\href@noop {} {\bibfield  {journal} {\bibinfo  {journal} {Science}\ }\textbf
  {\bibinfo {volume} {357}},\ \bibinfo {pages} {1369} (\bibinfo {year}
  {2017})}\BibitemShut {NoStop}%
\bibitem [{\citenamefont {Rowe}(1995)}]{Rowe4}%
  \BibitemOpen
  \bibfield  {author} {\bibinfo {author} {\bibfnamefont {D.~M.}\ \bibnamefont
  {Rowe}},\ }\href@noop {} {\bibfield  {journal} {\bibinfo  {journal} {CRC
  Press}\ }\textbf {\bibinfo {volume} {18}},\ \bibinfo {pages} {189} (\bibinfo
  {year} {1995})}\BibitemShut {NoStop}%
\bibitem [{\citenamefont {Wang}\ \emph {et~al.}(2013)\citenamefont {Wang},
  \citenamefont {Porter}, \citenamefont {B{\"o}ttner}, \citenamefont
  {K{\"o}nig}, \citenamefont {Chen}, \citenamefont {Bai}, \citenamefont
  {Tritt}, \citenamefont {Mayolet}, \citenamefont {Senawiratne}, \citenamefont
  {Smith}, \citenamefont {Harris}, \citenamefont {Girbert}, \citenamefont
  {Sharp}, \citenamefont {Lo}, \citenamefont {Kleinke},\ and\ \citenamefont
  {Kiss}}]{Wang5}%
  \BibitemOpen
  \bibfield  {author} {\bibinfo {author} {\bibfnamefont {H.}~\bibnamefont
  {Wang}}, \bibinfo {author} {\bibfnamefont {W.~D.}\ \bibnamefont {Porter}},
  \bibinfo {author} {\bibfnamefont {H.}~\bibnamefont {B{\"o}ttner}}, \bibinfo
  {author} {\bibfnamefont {J.}~\bibnamefont {K{\"o}nig}}, \bibinfo {author}
  {\bibfnamefont {L.}~\bibnamefont {Chen}}, \bibinfo {author} {\bibfnamefont
  {S.}~\bibnamefont {Bai}}, \bibinfo {author} {\bibfnamefont {T.~M.}\
  \bibnamefont {Tritt}}, \bibinfo {author} {\bibfnamefont {A.}~\bibnamefont
  {Mayolet}}, \bibinfo {author} {\bibfnamefont {J.}~\bibnamefont
  {Senawiratne}}, \bibinfo {author} {\bibfnamefont {C.}~\bibnamefont {Smith}},
  \bibinfo {author} {\bibfnamefont {F.}~\bibnamefont {Harris}}, \bibinfo
  {author} {\bibfnamefont {P.}~\bibnamefont {Girbert}}, \bibinfo {author}
  {\bibfnamefont {J.~W.}\ \bibnamefont {Sharp}}, \bibinfo {author}
  {\bibfnamefont {J.}~\bibnamefont {Lo}}, \bibinfo {author} {\bibfnamefont
  {H.}~\bibnamefont {Kleinke}}, \ and\ \bibinfo {author} {\bibfnamefont
  {L.}~\bibnamefont {Kiss}},\ }\href@noop {} {\bibfield  {journal} {\bibinfo
  {journal} {J. Electron. Mater.}\ }\textbf {\bibinfo {volume} {42}},\ \bibinfo
  {pages} {654} (\bibinfo {year} {2013})}\BibitemShut {NoStop}%
\bibitem [{\citenamefont {Wang}\ \emph {et~al.}(2015)\citenamefont {Wang},
  \citenamefont {Bai}, \citenamefont {Chen}, \citenamefont {Cuenat},
  \citenamefont {Joshi}, \citenamefont {Kleinke}, \citenamefont {K{\"o}nig},
  \citenamefont {Lee}, \citenamefont {Martin}, \citenamefont {Oh},
  \citenamefont {Porter}, \citenamefont {Ren}, \citenamefont {Salvador},
  \citenamefont {Sharp}, \citenamefont {Taylor}, \citenamefont {Thompson},\
  and\ \citenamefont {Tseng}}]{Wang6}%
  \BibitemOpen
  \bibfield  {author} {\bibinfo {author} {\bibfnamefont {H.}~\bibnamefont
  {Wang}}, \bibinfo {author} {\bibfnamefont {S.}~\bibnamefont {Bai}}, \bibinfo
  {author} {\bibfnamefont {L.}~\bibnamefont {Chen}}, \bibinfo {author}
  {\bibfnamefont {A.}~\bibnamefont {Cuenat}}, \bibinfo {author} {\bibfnamefont
  {G.}~\bibnamefont {Joshi}}, \bibinfo {author} {\bibfnamefont
  {H.}~\bibnamefont {Kleinke}}, \bibinfo {author} {\bibfnamefont
  {J.}~\bibnamefont {K{\"o}nig}}, \bibinfo {author} {\bibfnamefont {H.~W.}\
  \bibnamefont {Lee}}, \bibinfo {author} {\bibfnamefont {J.}~\bibnamefont
  {Martin}}, \bibinfo {author} {\bibfnamefont {M.-W.}\ \bibnamefont {Oh}},
  \bibinfo {author} {\bibfnamefont {W.~D.}\ \bibnamefont {Porter}}, \bibinfo
  {author} {\bibfnamefont {Z.}~\bibnamefont {Ren}}, \bibinfo {author}
  {\bibfnamefont {J.}~\bibnamefont {Salvador}}, \bibinfo {author}
  {\bibfnamefont {J.}~\bibnamefont {Sharp}}, \bibinfo {author} {\bibfnamefont
  {P.}~\bibnamefont {Taylor}}, \bibinfo {author} {\bibfnamefont {A.~J.}\
  \bibnamefont {Thompson}}, \ and\ \bibinfo {author} {\bibfnamefont {Y.~C.}\
  \bibnamefont {Tseng}},\ }\href@noop {} {\bibfield  {journal} {\bibinfo
  {journal} {J. Electron. Mater.}\ }\textbf {\bibinfo {volume} {44}},\ \bibinfo
  {pages} {4482} (\bibinfo {year} {2015})}\BibitemShut {NoStop}%
\bibitem [{\citenamefont {Alleno}\ \emph {et~al.}(2015)\citenamefont {Alleno},
  \citenamefont {B{\'e}rardan}, \citenamefont {Byl}, \citenamefont {Candolfi},
  \citenamefont {Daou}, \citenamefont {Decourt}, \citenamefont {Guilmeau},
  \citenamefont {H{\'e}bert}, \citenamefont {Hejtmanek}, \citenamefont
  {Lenoir}, \citenamefont {Masschelein}, \citenamefont {Ohorodnichuk},
  \citenamefont {Pollet}, \citenamefont {Populoh}, \citenamefont {Ravot},
  \citenamefont {Rouleau},\ and\ \citenamefont {Soulier}}]{AllenoCoNiSb}%
  \BibitemOpen
  \bibfield  {author} {\bibinfo {author} {\bibfnamefont {E.}~\bibnamefont
  {Alleno}}, \bibinfo {author} {\bibfnamefont {D.}~\bibnamefont
  {B{\'e}rardan}}, \bibinfo {author} {\bibfnamefont {C.}~\bibnamefont {Byl}},
  \bibinfo {author} {\bibfnamefont {C.}~\bibnamefont {Candolfi}}, \bibinfo
  {author} {\bibfnamefont {R.}~\bibnamefont {Daou}}, \bibinfo {author}
  {\bibfnamefont {R.}~\bibnamefont {Decourt}}, \bibinfo {author} {\bibfnamefont
  {E.}~\bibnamefont {Guilmeau}}, \bibinfo {author} {\bibfnamefont
  {S.}~\bibnamefont {H{\'e}bert}}, \bibinfo {author} {\bibfnamefont
  {J.}~\bibnamefont {Hejtmanek}}, \bibinfo {author} {\bibfnamefont
  {B.}~\bibnamefont {Lenoir}}, \bibinfo {author} {\bibfnamefont
  {P.}~\bibnamefont {Masschelein}}, \bibinfo {author} {\bibfnamefont
  {V.}~\bibnamefont {Ohorodnichuk}}, \bibinfo {author} {\bibfnamefont
  {M.}~\bibnamefont {Pollet}}, \bibinfo {author} {\bibfnamefont
  {S.}~\bibnamefont {Populoh}}, \bibinfo {author} {\bibfnamefont
  {D.}~\bibnamefont {Ravot}}, \bibinfo {author} {\bibfnamefont
  {O.}~\bibnamefont {Rouleau}}, \ and\ \bibinfo {author} {\bibfnamefont
  {M.}~\bibnamefont {Soulier}},\ }\href@noop {} {\bibfield  {journal} {\bibinfo
   {journal} {Rev. Sci. Instrum.}\ }\textbf {\bibinfo {volume} {86}},\ \bibinfo
  {pages} {011301} (\bibinfo {year} {2015})}\BibitemShut {NoStop}%
\bibitem [{\citenamefont {Heremans}\ and\ \citenamefont
  {Martin}(2024)}]{HeremansNM}%
  \BibitemOpen
  \bibfield  {author} {\bibinfo {author} {\bibfnamefont {J.~P.}\ \bibnamefont
  {Heremans}}\ and\ \bibinfo {author} {\bibfnamefont {J.}~\bibnamefont
  {Martin}},\ }\href@noop {} {\bibfield  {journal} {\bibinfo  {journal} {Nature
  Materials}\ }\textbf {\bibinfo {volume} {23}},\ \bibinfo {pages} {18}
  (\bibinfo {year} {2024})}\BibitemShut {NoStop}%
\bibitem [{\citenamefont {Harman}(1958)}]{Harman7}%
  \BibitemOpen
  \bibfield  {author} {\bibinfo {author} {\bibfnamefont {T.~C.}\ \bibnamefont
  {Harman}},\ }\href@noop {} {\bibfield  {journal} {\bibinfo  {journal} {J.
  Appl. Phys.}\ }\textbf {\bibinfo {volume} {29}},\ \bibinfo {pages} {1373}
  (\bibinfo {year} {1958})}\BibitemShut {NoStop}%
\bibitem [{\citenamefont {Satake}\ \emph {et~al.}(2004)\citenamefont {Satake},
  \citenamefont {Tanaka}, \citenamefont {Ohkawa}, \citenamefont {Fujii},\ and\
  \citenamefont {Terasaki}}]{Satake8}%
  \BibitemOpen
  \bibfield  {author} {\bibinfo {author} {\bibfnamefont {A.}~\bibnamefont
  {Satake}}, \bibinfo {author} {\bibfnamefont {H.}~\bibnamefont {Tanaka}},
  \bibinfo {author} {\bibfnamefont {T.}~\bibnamefont {Ohkawa}}, \bibinfo
  {author} {\bibfnamefont {T.}~\bibnamefont {Fujii}}, \ and\ \bibinfo {author}
  {\bibfnamefont {I.}~\bibnamefont {Terasaki}},\ }\href@noop {} {\bibfield
  {journal} {\bibinfo  {journal} {J. Appl. Phys.}\ }\textbf {\bibinfo {volume}
  {96}},\ \bibinfo {pages} {931} (\bibinfo {year} {2004})}\BibitemShut
  {NoStop}%
\bibitem [{\citenamefont {Kobayashi}\ \emph {et~al.}(2008)\citenamefont
  {Kobayashi}, \citenamefont {Tamura},\ and\ \citenamefont
  {Terasaki}}]{Kobayashi9}%
  \BibitemOpen
  \bibfield  {author} {\bibinfo {author} {\bibfnamefont {W.}~\bibnamefont
  {Kobayashi}}, \bibinfo {author} {\bibfnamefont {W.}~\bibnamefont {Tamura}}, \
  and\ \bibinfo {author} {\bibfnamefont {I.}~\bibnamefont {Terasaki}},\
  }\href@noop {} {\bibfield  {journal} {\bibinfo  {journal} {J. Phys. Soc.
  Jpn.}\ }\textbf {\bibinfo {volume} {77}},\ \bibinfo {pages} {07606} (\bibinfo
  {year} {2008})}\BibitemShut {NoStop}%
\bibitem [{\citenamefont {Kobayashi}\ \emph {et~al.}(2009)\citenamefont
  {Kobayashi}, \citenamefont {Tamura},\ and\ \citenamefont
  {Terasaki}}]{Kobayashi10}%
  \BibitemOpen
  \bibfield  {author} {\bibinfo {author} {\bibfnamefont {W.}~\bibnamefont
  {Kobayashi}}, \bibinfo {author} {\bibfnamefont {W.}~\bibnamefont {Tamura}}, \
  and\ \bibinfo {author} {\bibfnamefont {I.}~\bibnamefont {Terasaki}},\
  }\href@noop {} {\bibfield  {journal} {\bibinfo  {journal} {J. Electron.
  Mater.}\ }\textbf {\bibinfo {volume} {38}},\ \bibinfo {pages} {964} (\bibinfo
  {year} {2009})}\BibitemShut {NoStop}%
\bibitem [{\citenamefont {Singh}\ \emph {et~al.}(2009)\citenamefont {Singh},
  \citenamefont {Bian}, \citenamefont {Shakouri}, \citenamefont {Zeng},
  \citenamefont {Bahk}, \citenamefont {Bowers}, \citenamefont {Zide},\ and\
  \citenamefont {Gossard}}]{Singh11}%
  \BibitemOpen
  \bibfield  {author} {\bibinfo {author} {\bibfnamefont {R.}~\bibnamefont
  {Singh}}, \bibinfo {author} {\bibfnamefont {Z.}~\bibnamefont {Bian}},
  \bibinfo {author} {\bibfnamefont {A.}~\bibnamefont {Shakouri}}, \bibinfo
  {author} {\bibfnamefont {G.}~\bibnamefont {Zeng}}, \bibinfo {author}
  {\bibfnamefont {J.-H.}\ \bibnamefont {Bahk}}, \bibinfo {author}
  {\bibfnamefont {J.~E.}\ \bibnamefont {Bowers}}, \bibinfo {author}
  {\bibfnamefont {J.~M.~O.}\ \bibnamefont {Zide}}, \ and\ \bibinfo {author}
  {\bibfnamefont {A.~C.}\ \bibnamefont {Gossard}},\ }\href@noop {} {\bibfield
  {journal} {\bibinfo  {journal} {Appl. Phys. Lett.}\ }\textbf {\bibinfo
  {volume} {94}},\ \bibinfo {pages} {212508} (\bibinfo {year}
  {2009})}\BibitemShut {NoStop}%
\bibitem [{\citenamefont {Venkatasubramanlan}\ \emph
  {et~al.}(2001)\citenamefont {Venkatasubramanlan}, \citenamefont {Silvola},
  \citenamefont {Colpitts},\ and\ \citenamefont
  {O'Quinn}}]{Venkatasubramanlan12}%
  \BibitemOpen
  \bibfield  {author} {\bibinfo {author} {\bibfnamefont {R.}~\bibnamefont
  {Venkatasubramanlan}}, \bibinfo {author} {\bibfnamefont {E.}~\bibnamefont
  {Silvola}}, \bibinfo {author} {\bibfnamefont {T.}~\bibnamefont {Colpitts}}, \
  and\ \bibinfo {author} {\bibfnamefont {B.}~\bibnamefont {O'Quinn}},\
  }\href@noop {} {\bibfield  {journal} {\bibinfo  {journal} {Nature}\ }\textbf
  {\bibinfo {volume} {413}},\ \bibinfo {pages} {597} (\bibinfo {year}
  {2001})}\BibitemShut {NoStop}%
\bibitem [{\citenamefont {Iwasaki}\ \emph {et~al.}(2003)\citenamefont
  {Iwasaki}, \citenamefont {Yokoyama}, \citenamefont {Tsukui}, \citenamefont
  {Koyano}, \citenamefont {Hori},\ and\ \citenamefont {Sano}}]{Iwasaki13}%
  \BibitemOpen
  \bibfield  {author} {\bibinfo {author} {\bibfnamefont {H.}~\bibnamefont
  {Iwasaki}}, \bibinfo {author} {\bibfnamefont {S.}~\bibnamefont {Yokoyama}},
  \bibinfo {author} {\bibfnamefont {T.}~\bibnamefont {Tsukui}}, \bibinfo
  {author} {\bibfnamefont {M.}~\bibnamefont {Koyano}}, \bibinfo {author}
  {\bibfnamefont {H.}~\bibnamefont {Hori}}, \ and\ \bibinfo {author}
  {\bibfnamefont {S.}~\bibnamefont {Sano}},\ }\href@noop {} {\bibfield
  {journal} {\bibinfo  {journal} {Jpn. J. Appl. Phys.}\ }\textbf {\bibinfo
  {volume} {42}},\ \bibinfo {pages} {3703} (\bibinfo {year}
  {2003})}\BibitemShut {NoStop}%
\bibitem [{\citenamefont {Korzhuev}\ \emph {et~al.}(2011)\citenamefont
  {Korzhuev}, \citenamefont {Avilov},\ and\ \citenamefont
  {Nichezina}}]{Korzhuev47}%
  \BibitemOpen
  \bibfield  {author} {\bibinfo {author} {\bibfnamefont {M.~A.}\ \bibnamefont
  {Korzhuev}}, \bibinfo {author} {\bibfnamefont {E.~S.}\ \bibnamefont
  {Avilov}}, \ and\ \bibinfo {author} {\bibfnamefont {I.~Y.}\ \bibnamefont
  {Nichezina}},\ }\href@noop {} {\bibfield  {journal} {\bibinfo  {journal} {J.
  Electron. Mater.}\ }\textbf {\bibinfo {volume} {40}},\ \bibinfo {pages} {733}
  (\bibinfo {year} {2011})}\BibitemShut {NoStop}%
\bibitem [{\citenamefont {Barako}\ \emph {et~al.}(2012)\citenamefont {Barako},
  \citenamefont {Park}, \citenamefont {Marconnet}, \citenamefont {M.Asheghi},\
  and\ \citenamefont {Goodson}}]{Barako48}%
  \BibitemOpen
  \bibfield  {author} {\bibinfo {author} {\bibfnamefont {M.}~\bibnamefont
  {Barako}}, \bibinfo {author} {\bibfnamefont {W.}~\bibnamefont {Park}},
  \bibinfo {author} {\bibfnamefont {A.}~\bibnamefont {Marconnet}}, \bibinfo
  {author} {\bibnamefont {M.Asheghi}}, \ and\ \bibinfo {author} {\bibfnamefont
  {K.}~\bibnamefont {Goodson}},\ }\href@noop {} {\bibfield  {journal} {\bibinfo
   {journal} {J. Electron. Mater.}\ }\textbf {\bibinfo {volume} {42}},\
  \bibinfo {pages} {372} (\bibinfo {year} {2012})}\BibitemShut {NoStop}%
\bibitem [{\citenamefont {Muto}\ \emph {et~al.}(2009)\citenamefont {Muto},
  \citenamefont {Kraemer}, \citenamefont {Hao}, \citenamefont {Ren},\ and\
  \citenamefont {Chen}}]{Muto49}%
  \BibitemOpen
  \bibfield  {author} {\bibinfo {author} {\bibfnamefont {A.}~\bibnamefont
  {Muto}}, \bibinfo {author} {\bibfnamefont {D.}~\bibnamefont {Kraemer}},
  \bibinfo {author} {\bibfnamefont {Q.}~\bibnamefont {Hao}}, \bibinfo {author}
  {\bibfnamefont {Z.~F.}\ \bibnamefont {Ren}}, \ and\ \bibinfo {author}
  {\bibfnamefont {G.}~\bibnamefont {Chen}},\ }\href@noop {} {\bibfield
  {journal} {\bibinfo  {journal} {Rev. Sci. Instrum.}\ }\textbf {\bibinfo
  {volume} {80}},\ \bibinfo {pages} {093901} (\bibinfo {year}
  {2009})}\BibitemShut {NoStop}%
\bibitem [{\citenamefont {Iwasaki}\ \emph {et~al.}(2004)\citenamefont
  {Iwasaki}, \citenamefont {Koyano}, \citenamefont {Yamamura},\ and\
  \citenamefont {Hori}}]{Iwasaki14}%
  \BibitemOpen
  \bibfield  {author} {\bibinfo {author} {\bibfnamefont {H.}~\bibnamefont
  {Iwasaki}}, \bibinfo {author} {\bibfnamefont {M.}~\bibnamefont {Koyano}},
  \bibinfo {author} {\bibfnamefont {Y.}~\bibnamefont {Yamamura}}, \ and\
  \bibinfo {author} {\bibfnamefont {H.}~\bibnamefont {Hori}},\ }\href@noop {}
  {\bibfield  {journal} {\bibinfo  {journal} {Solid State Commun.}\ }\textbf
  {\bibinfo {volume} {130}},\ \bibinfo {pages} {507} (\bibinfo {year}
  {2004})}\BibitemShut {NoStop}%
\bibitem [{\citenamefont {Iwasaki}\ and\ \citenamefont
  {Hori}(2015)}]{Iwasaki15}%
  \BibitemOpen
  \bibfield  {author} {\bibinfo {author} {\bibfnamefont {H.}~\bibnamefont
  {Iwasaki}}\ and\ \bibinfo {author} {\bibfnamefont {H.}~\bibnamefont {Hori}},\
  }\href@noop {} {\bibfield  {journal} {\bibinfo  {journal} {in International
  Conference on Thermoelectrics, ICT Proceedings}\ ,\ \bibinfo {pages} {501}}
  (\bibinfo {year} {2015})}\BibitemShut {NoStop}%
\bibitem [{\citenamefont {Downey}\ \emph {et~al.}(2007)\citenamefont {Downey},
  \citenamefont {Hogan},\ and\ \citenamefont {Cook}}]{Downey16}%
  \BibitemOpen
  \bibfield  {author} {\bibinfo {author} {\bibfnamefont {A.~D.}\ \bibnamefont
  {Downey}}, \bibinfo {author} {\bibfnamefont {T.~P.}\ \bibnamefont {Hogan}}, \
  and\ \bibinfo {author} {\bibfnamefont {B.}~\bibnamefont {Cook}},\ }\href@noop
  {} {\bibfield  {journal} {\bibinfo  {journal} {Rev. Sci. Instrum.}\ }\textbf
  {\bibinfo {volume} {78}},\ \bibinfo {pages} {093904} (\bibinfo {year}
  {2007})}\BibitemShut {NoStop}%
\bibitem [{\citenamefont {Okawa}\ \emph {et~al.}(2021)\citenamefont {Okawa},
  \citenamefont {Amagai}, \citenamefont {Fujiki},\ and\ \citenamefont
  {Kaneko}}]{Okawa17}%
  \BibitemOpen
  \bibfield  {author} {\bibinfo {author} {\bibfnamefont {K.}~\bibnamefont
  {Okawa}}, \bibinfo {author} {\bibfnamefont {Y.}~\bibnamefont {Amagai}},
  \bibinfo {author} {\bibfnamefont {H.}~\bibnamefont {Fujiki}}, \ and\ \bibinfo
  {author} {\bibfnamefont {N.-H.}\ \bibnamefont {Kaneko}},\ }\href@noop {}
  {\bibfield  {journal} {\bibinfo  {journal} {Commun. Phys.}\ }\textbf
  {\bibinfo {volume} {4}},\ \bibinfo {pages} {267} (\bibinfo {year}
  {2021})}\BibitemShut {NoStop}%
\bibitem [{\citenamefont {Okawa}\ \emph {et~al.}(2020)\citenamefont {Okawa},
  \citenamefont {Amagai}, \citenamefont {Fujiki}, \citenamefont {Kaneko},
  \citenamefont {Tsuchimine}, \citenamefont {Kaneko}, \citenamefont {Tasaki},
  \citenamefont {Ohata}, \citenamefont {Okajima},\ and\ \citenamefont
  {Nambu}}]{OkawaJJAP}%
  \BibitemOpen
  \bibfield  {author} {\bibinfo {author} {\bibfnamefont {K.}~\bibnamefont
  {Okawa}}, \bibinfo {author} {\bibfnamefont {Y.}~\bibnamefont {Amagai}},
  \bibinfo {author} {\bibfnamefont {H.}~\bibnamefont {Fujiki}}, \bibinfo
  {author} {\bibfnamefont {N.-H.}\ \bibnamefont {Kaneko}}, \bibinfo {author}
  {\bibfnamefont {N.}~\bibnamefont {Tsuchimine}}, \bibinfo {author}
  {\bibfnamefont {H.}~\bibnamefont {Kaneko}}, \bibinfo {author} {\bibfnamefont
  {Y.}~\bibnamefont {Tasaki}}, \bibinfo {author} {\bibfnamefont
  {K.}~\bibnamefont {Ohata}}, \bibinfo {author} {\bibfnamefont
  {M.}~\bibnamefont {Okajima}}, \ and\ \bibinfo {author} {\bibfnamefont
  {S.}~\bibnamefont {Nambu}},\ }\href@noop {} {\bibfield  {journal} {\bibinfo
  {journal} {Japanese Journal of Applied Physics}\ }\textbf {\bibinfo {volume}
  {59}},\ \bibinfo {pages} {046504} (\bibinfo {year} {2020})}\BibitemShut
  {NoStop}%
\bibitem [{\citenamefont {Ao}\ \emph {et~al.}(2011)\citenamefont {Ao},
  \citenamefont {de~Boor},\ and\ \citenamefont {Schmidt}}]{Ao18}%
  \BibitemOpen
  \bibfield  {author} {\bibinfo {author} {\bibfnamefont {X.}~\bibnamefont
  {Ao}}, \bibinfo {author} {\bibfnamefont {J.}~\bibnamefont {de~Boor}}, \ and\
  \bibinfo {author} {\bibfnamefont {V.}~\bibnamefont {Schmidt}},\ }\href@noop
  {} {\bibfield  {journal} {\bibinfo  {journal} {Adv. Energy Mater.}\ }\textbf
  {\bibinfo {volume} {1}},\ \bibinfo {pages} {1007} (\bibinfo {year}
  {2011})}\BibitemShut {NoStop}%
\bibitem [{\citenamefont {Shinozaki}\ \emph {et~al.}(2020)\citenamefont
  {Shinozaki}, \citenamefont {Hirabayashi},\ and\ \citenamefont
  {Hasegawa}}]{Shinozaki19}%
  \BibitemOpen
  \bibfield  {author} {\bibinfo {author} {\bibfnamefont {R.}~\bibnamefont
  {Shinozaki}}, \bibinfo {author} {\bibfnamefont {S.}~\bibnamefont
  {Hirabayashi}}, \ and\ \bibinfo {author} {\bibfnamefont {Y.}~\bibnamefont
  {Hasegawa}},\ }\href@noop {} {\bibfield  {journal} {\bibinfo  {journal}
  {Appl. Phys. Express}\ }\textbf {\bibinfo {volume} {13}},\ \bibinfo {pages}
  {106501} (\bibinfo {year} {2020})}\BibitemShut {NoStop}%
\bibitem [{\citenamefont {Hasegawa}\ and\ \citenamefont
  {Takeuchi}(2022)}]{Hasegawa20}%
  \BibitemOpen
  \bibfield  {author} {\bibinfo {author} {\bibfnamefont {Y.}~\bibnamefont
  {Hasegawa}}\ and\ \bibinfo {author} {\bibfnamefont {M.}~\bibnamefont
  {Takeuchi}},\ }\href@noop {} {\bibfield  {journal} {\bibinfo  {journal} {Sci.
  Rep.}\ }\textbf {\bibinfo {volume} {12}},\ \bibinfo {pages} {11967} (\bibinfo
  {year} {2022})}\BibitemShut {NoStop}%
\bibitem [{\citenamefont {Hasegawa}\ and\ \citenamefont
  {Takeuchi}(2023)}]{Hasegawa21}%
  \BibitemOpen
  \bibfield  {author} {\bibinfo {author} {\bibfnamefont {Y.}~\bibnamefont
  {Hasegawa}}\ and\ \bibinfo {author} {\bibfnamefont {M.}~\bibnamefont
  {Takeuchi}},\ }\href@noop {} {\bibfield  {journal} {\bibinfo  {journal} {Rev.
  Sci. Instrum.}\ }\textbf {\bibinfo {volume} {94}},\ \bibinfo {pages} {014902}
  (\bibinfo {year} {2023})}\BibitemShut {NoStop}%
\bibitem [{\citenamefont {Marchi}\ and\ \citenamefont
  {Giaretto}(2011)}]{Marchi22}%
  \BibitemOpen
  \bibfield  {author} {\bibinfo {author} {\bibfnamefont {A.~D.}\ \bibnamefont
  {Marchi}}\ and\ \bibinfo {author} {\bibfnamefont {V.}~\bibnamefont
  {Giaretto}},\ }\href@noop {} {\bibfield  {journal} {\bibinfo  {journal} {Rev.
  Sci. Instrum.}\ }\textbf {\bibinfo {volume} {82}},\ \bibinfo {pages} {034901}
  (\bibinfo {year} {2011})}\BibitemShut {NoStop}%
\bibitem [{\citenamefont {Marchi}\ \emph {et~al.}(2013)\citenamefont {Marchi},
  \citenamefont {Giaretto}, \citenamefont {Caron},\ and\ \citenamefont
  {Tona}}]{Marchi23}%
  \BibitemOpen
  \bibfield  {author} {\bibinfo {author} {\bibfnamefont {A.~D.}\ \bibnamefont
  {Marchi}}, \bibinfo {author} {\bibfnamefont {V.}~\bibnamefont {Giaretto}},
  \bibinfo {author} {\bibfnamefont {S.}~\bibnamefont {Caron}}, \ and\ \bibinfo
  {author} {\bibfnamefont {A.}~\bibnamefont {Tona}},\ }\href@noop {} {\bibfield
   {journal} {\bibinfo  {journal} {J. Electron. Mater.}\ }\textbf {\bibinfo
  {volume} {42}},\ \bibinfo {pages} {2067} (\bibinfo {year}
  {2013})}\BibitemShut {NoStop}%
\bibitem [{\citenamefont {Apertet}\ and\ \citenamefont
  {Ouerdane}(2017)}]{Apertet24}%
  \BibitemOpen
  \bibfield  {author} {\bibinfo {author} {\bibfnamefont {Y.}~\bibnamefont
  {Apertet}}\ and\ \bibinfo {author} {\bibfnamefont {H.}~\bibnamefont
  {Ouerdane}},\ }\href@noop {} {\bibfield  {journal} {\bibinfo  {journal}
  {Energy Conversion and Management}\ }\textbf {\bibinfo {volume} {149}},\
  \bibinfo {pages} {564} (\bibinfo {year} {2017})}\BibitemShut {NoStop}%
\bibitem [{\citenamefont {Beltran-Pitarch}\ and\ \citenamefont
  {Prado-Gonjal}(2018)}]{Pitarch25}%
  \BibitemOpen
  \bibfield  {author} {\bibinfo {author} {\bibfnamefont {B.}~\bibnamefont
  {Beltran-Pitarch}}\ and\ \bibinfo {author} {\bibfnamefont {J.}~\bibnamefont
  {Prado-Gonjal}},\ }\href@noop {} {\bibfield  {journal} {\bibinfo  {journal}
  {J. Appl. Phys.}\ }\textbf {\bibinfo {volume} {123}},\ \bibinfo {pages}
  {084505} (\bibinfo {year} {2018})}\BibitemShut {NoStop}%
\bibitem [{\citenamefont {Beltran-Pitarch}\ \emph
  {et~al.}(2020{\natexlab{a}})\citenamefont {Beltran-Pitarch}, \citenamefont
  {Prado-Gonjal}, \citenamefont {Powell}, \citenamefont {Ziolkowski},\ and\
  \citenamefont {Garcia-Canadas}}]{Pitarch26}%
  \BibitemOpen
  \bibfield  {author} {\bibinfo {author} {\bibfnamefont {B.}~\bibnamefont
  {Beltran-Pitarch}}, \bibinfo {author} {\bibfnamefont {J.}~\bibnamefont
  {Prado-Gonjal}}, \bibinfo {author} {\bibfnamefont {A.~V.}\ \bibnamefont
  {Powell}}, \bibinfo {author} {\bibfnamefont {P.}~\bibnamefont {Ziolkowski}},
  \ and\ \bibinfo {author} {\bibfnamefont {J.}~\bibnamefont {Garcia-Canadas}},\
  }\href@noop {} {\bibfield  {journal} {\bibinfo  {journal} {J. Appl. Phys.}\
  }\textbf {\bibinfo {volume} {165}},\ \bibinfo {pages} {114361} (\bibinfo
  {year} {2020}{\natexlab{a}})}\BibitemShut {NoStop}%
\bibitem [{\citenamefont {Arisaka}\ \emph {et~al.}(2019)\citenamefont
  {Arisaka}, \citenamefont {Otsuka},\ and\ \citenamefont
  {Hasegawa}}]{Arisaka27}%
  \BibitemOpen
  \bibfield  {author} {\bibinfo {author} {\bibfnamefont {T.}~\bibnamefont
  {Arisaka}}, \bibinfo {author} {\bibfnamefont {M.}~\bibnamefont {Otsuka}}, \
  and\ \bibinfo {author} {\bibfnamefont {Y.}~\bibnamefont {Hasegawa}},\
  }\href@noop {} {\bibfield  {journal} {\bibinfo  {journal} {Rev. Sci.
  Instrum.}\ }\textbf {\bibinfo {volume} {90}},\ \bibinfo {pages} {046104}
  (\bibinfo {year} {2019})}\BibitemShut {NoStop}%
\bibitem [{\citenamefont {Beltran-Pitarch}\ \emph
  {et~al.}(2020{\natexlab{b}})\citenamefont {Beltran-Pitarch}, \citenamefont
  {Vidan},\ and\ \citenamefont {Garcia-Canadas}}]{Pitarch28}%
  \BibitemOpen
  \bibfield  {author} {\bibinfo {author} {\bibfnamefont {B.}~\bibnamefont
  {Beltran-Pitarch}}, \bibinfo {author} {\bibfnamefont {F.}~\bibnamefont
  {Vidan}}, \ and\ \bibinfo {author} {\bibfnamefont {J.}~\bibnamefont
  {Garcia-Canadas}},\ }\href@noop {} {\bibfield  {journal} {\bibinfo  {journal}
  {Appl. Thermal Engineering}\ }\textbf {\bibinfo {volume} {165}},\ \bibinfo
  {pages} {114361} (\bibinfo {year} {2020}{\natexlab{b}})}\BibitemShut
  {NoStop}%
\bibitem [{\citenamefont {Yoo}\ \emph {et~al.}(2019)\citenamefont {Yoo},
  \citenamefont {Yeon}, \citenamefont {Jin}, \citenamefont {Kim}, \citenamefont
  {Song}, \citenamefont {Yoon}, \citenamefont {Park}, \citenamefont
  {Beltran-Pitarch}, \citenamefont {Prado-Gonjal},\ and\ \citenamefont
  {Min}}]{Yoo29}%
  \BibitemOpen
  \bibfield  {author} {\bibinfo {author} {\bibfnamefont {C.-Y.}\ \bibnamefont
  {Yoo}}, \bibinfo {author} {\bibfnamefont {C.}~\bibnamefont {Yeon}}, \bibinfo
  {author} {\bibfnamefont {Y.}~\bibnamefont {Jin}}, \bibinfo {author}
  {\bibfnamefont {Y.}~\bibnamefont {Kim}}, \bibinfo {author} {\bibfnamefont
  {J.}~\bibnamefont {Song}}, \bibinfo {author} {\bibfnamefont {H.}~\bibnamefont
  {Yoon}}, \bibinfo {author} {\bibfnamefont {S.-H.}\ \bibnamefont {Park}},
  \bibinfo {author} {\bibfnamefont {B.}~\bibnamefont {Beltran-Pitarch}},
  \bibinfo {author} {\bibfnamefont {J.}~\bibnamefont {Prado-Gonjal}}, \ and\
  \bibinfo {author} {\bibfnamefont {G.}~\bibnamefont {Min}},\ }\href@noop {}
  {\bibfield  {journal} {\bibinfo  {journal} {Appl. Energy}\ }\textbf {\bibinfo
  {volume} {251}},\ \bibinfo {pages} {113341} (\bibinfo {year}
  {2019})}\BibitemShut {NoStop}%
\bibitem [{\citenamefont {Otsuka}\ \emph {et~al.}(2020)\citenamefont {Otsuka},
  \citenamefont {Arisaka},\ and\ \citenamefont {Hasegawa}}]{Otsuka30}%
  \BibitemOpen
  \bibfield  {author} {\bibinfo {author} {\bibfnamefont {M.}~\bibnamefont
  {Otsuka}}, \bibinfo {author} {\bibfnamefont {T.}~\bibnamefont {Arisaka}}, \
  and\ \bibinfo {author} {\bibfnamefont {Y.}~\bibnamefont {Hasegawa}},\
  }\href@noop {} {\bibfield  {journal} {\bibinfo  {journal} {Materials Science
  \& Engineering B}\ }\textbf {\bibinfo {volume} {261}},\ \bibinfo {pages}
  {114620} (\bibinfo {year} {2020})}\BibitemShut {NoStop}%
\bibitem [{\citenamefont {Zaoui}\ \emph {et~al.}(2020)\citenamefont {Zaoui},
  \citenamefont {Belayadi}, \citenamefont {M.Zabat}, \citenamefont {Mougari},\
  and\ \citenamefont {Mekideche-Chafa}}]{Zaoui31}%
  \BibitemOpen
  \bibfield  {author} {\bibinfo {author} {\bibfnamefont {S.}~\bibnamefont
  {Zaoui}}, \bibinfo {author} {\bibfnamefont {A.}~\bibnamefont {Belayadi}},
  \bibinfo {author} {\bibnamefont {M.Zabat}}, \bibinfo {author} {\bibfnamefont
  {A.}~\bibnamefont {Mougari}}, \ and\ \bibinfo {author} {\bibfnamefont
  {F.}~\bibnamefont {Mekideche-Chafa}},\ }\href@noop {} {\bibfield  {journal}
  {\bibinfo  {journal} {Physica B}\ }\textbf {\bibinfo {volume} {580}},\
  \bibinfo {pages} {411735} (\bibinfo {year} {2020})}\BibitemShut {NoStop}%
\bibitem [{\citenamefont {Hasegawa}\ and\ \citenamefont
  {Takeuchi}(2021)}]{Hasegawa32}%
  \BibitemOpen
  \bibfield  {author} {\bibinfo {author} {\bibfnamefont {Y.}~\bibnamefont
  {Hasegawa}}\ and\ \bibinfo {author} {\bibfnamefont {M.}~\bibnamefont
  {Takeuchi}},\ }\href@noop {} {\bibfield  {journal} {\bibinfo  {journal} {Rev.
  Sci. Instrum.}\ }\textbf {\bibinfo {volume} {92}},\ \bibinfo {pages} {083902}
  (\bibinfo {year} {2021})}\BibitemShut {NoStop}%
\bibitem [{\citenamefont {Beltran-Pitarch}\ \emph {et~al.}(2021)\citenamefont
  {Beltran-Pitarch}, \citenamefont {Maassen},\ and\ \citenamefont
  {Garcia-Canadas}}]{Pitarch33}%
  \BibitemOpen
  \bibfield  {author} {\bibinfo {author} {\bibfnamefont {B.}~\bibnamefont
  {Beltran-Pitarch}}, \bibinfo {author} {\bibfnamefont {J.}~\bibnamefont
  {Maassen}}, \ and\ \bibinfo {author} {\bibfnamefont {J.}~\bibnamefont
  {Garcia-Canadas}},\ }\href@noop {} {\bibfield  {journal} {\bibinfo  {journal}
  {Appl. Energy}\ }\textbf {\bibinfo {volume} {299}},\ \bibinfo {pages}
  {117287} (\bibinfo {year} {2021})}\BibitemShut {NoStop}%
\bibitem [{\citenamefont {Hirabayashi}\ and\ \citenamefont
  {Hasegawa}(2021)}]{Hirabayashi34}%
  \BibitemOpen
  \bibfield  {author} {\bibinfo {author} {\bibfnamefont {S.}~\bibnamefont
  {Hirabayashi}}\ and\ \bibinfo {author} {\bibfnamefont {Y.}~\bibnamefont
  {Hasegawa}},\ }\href@noop {} {\bibfield  {journal} {\bibinfo  {journal} {Jpn.
  J. Appl. Phys.}\ }\textbf {\bibinfo {volume} {60}},\ \bibinfo {pages}
  {106503} (\bibinfo {year} {2021})}\BibitemShut {NoStop}%
\bibitem [{\citenamefont {Korzhuev}\ and\ \citenamefont
  {Avilov}(2010)}]{Korzhuev35}%
  \BibitemOpen
  \bibfield  {author} {\bibinfo {author} {\bibfnamefont {M.~A.}\ \bibnamefont
  {Korzhuev}}\ and\ \bibinfo {author} {\bibfnamefont {E.~S.}\ \bibnamefont
  {Avilov}},\ }\href@noop {} {\bibfield  {journal} {\bibinfo  {journal} {J.
  Electron. Mater.}\ }\textbf {\bibinfo {volume} {39}},\ \bibinfo {pages}
  {1499} (\bibinfo {year} {2010})}\BibitemShut {NoStop}%
\bibitem [{\citenamefont {Chen}(2011)}]{ChenCp}%
  \BibitemOpen
  \bibfield  {author} {\bibinfo {author} {\bibfnamefont {X.}~\bibnamefont
  {Chen}},\ }\href@noop {} {\bibfield  {journal} {\bibinfo  {journal} {Appl.
  Phys. Lett.}\ }\textbf {\bibinfo {volume} {99}},\ \bibinfo {pages} {261912}
  (\bibinfo {year} {2011})}\BibitemShut {NoStop}%
\bibitem [{\citenamefont {d.~Boor}\ and\ \citenamefont
  {Schmidt}(2010)}]{BoorVDP}%
  \BibitemOpen
  \bibfield  {author} {\bibinfo {author} {\bibfnamefont {J.}~\bibnamefont
  {d.~Boor}}\ and\ \bibinfo {author} {\bibfnamefont {V.}~\bibnamefont
  {Schmidt}},\ }\href@noop {} {\bibfield  {journal} {\bibinfo  {journal} {Adv.
  Mater.}\ }\textbf {\bibinfo {volume} {22}},\ \bibinfo {pages} {4303}
  (\bibinfo {year} {2010})}\BibitemShut {NoStop}%
\bibitem [{\citenamefont {Kwon}\ \emph {et~al.}(2014)\citenamefont {Kwon},
  \citenamefont {Baek}, \citenamefont {Kim},\ and\ \citenamefont
  {Kim}}]{Kwon37}%
  \BibitemOpen
  \bibfield  {author} {\bibinfo {author} {\bibfnamefont {B.}~\bibnamefont
  {Kwon}}, \bibinfo {author} {\bibfnamefont {S.-H.}\ \bibnamefont {Baek}},
  \bibinfo {author} {\bibfnamefont {S.~K.}\ \bibnamefont {Kim}}, \ and\
  \bibinfo {author} {\bibfnamefont {J.-S.}\ \bibnamefont {Kim}},\ }\href@noop
  {} {\bibfield  {journal} {\bibinfo  {journal} {Rev. Sci. Instrum.}\ }\textbf
  {\bibinfo {volume} {85}},\ \bibinfo {pages} {045108} (\bibinfo {year}
  {2014})}\BibitemShut {NoStop}%
\bibitem [{\citenamefont {Penn}(1964)}]{Penn38}%
  \BibitemOpen
  \bibfield  {author} {\bibinfo {author} {\bibfnamefont {A.~W.}\ \bibnamefont
  {Penn}},\ }\href@noop {} {\bibfield  {journal} {\bibinfo  {journal} {J. Sci.
  Instrum}\ }\textbf {\bibinfo {volume} {41}},\ \bibinfo {pages} {626}
  (\bibinfo {year} {1964})}\BibitemShut {NoStop}%
\bibitem [{\citenamefont {Campbell}\ \emph {et~al.}(1965)\citenamefont
  {Campbell}, \citenamefont {Hogarth},\ and\ \citenamefont
  {Hagger}}]{Campbell39}%
  \BibitemOpen
  \bibfield  {author} {\bibinfo {author} {\bibfnamefont {M.~R.}\ \bibnamefont
  {Campbell}}, \bibinfo {author} {\bibfnamefont {C.~A.}\ \bibnamefont
  {Hogarth}}, \ and\ \bibinfo {author} {\bibfnamefont {C.~A.}\ \bibnamefont
  {Hagger}},\ }\href@noop {} {\bibfield  {journal} {\bibinfo  {journal} {Int.
  J. Electron.}\ }\textbf {\bibinfo {volume} {19}},\ \bibinfo {pages} {571}
  (\bibinfo {year} {1965})}\BibitemShut {NoStop}%
\bibitem [{\citenamefont {Iwasaki}\ \emph {et~al.}(2002)\citenamefont
  {Iwasaki}, \citenamefont {Koyano},\ and\ \citenamefont {H.Hori}}]{Iwasaki40}%
  \BibitemOpen
  \bibfield  {author} {\bibinfo {author} {\bibfnamefont {H.}~\bibnamefont
  {Iwasaki}}, \bibinfo {author} {\bibfnamefont {M.}~\bibnamefont {Koyano}}, \
  and\ \bibinfo {author} {\bibnamefont {H.Hori}},\ }\href@noop {} {\bibfield
  {journal} {\bibinfo  {journal} {Jpn. J. Appl. Phys.}\ }\textbf {\bibinfo
  {volume} {41}},\ \bibinfo {pages} {6606} (\bibinfo {year}
  {2002})}\BibitemShut {NoStop}%
\bibitem [{\citenamefont {Kang}\ \emph {et~al.}(2016)\citenamefont {Kang},
  \citenamefont {Roh}, \citenamefont {Lee}, \citenamefont {Baek}, \citenamefont
  {Kim}, \citenamefont {Ju}, \citenamefont {Hyun}, \citenamefont {Kim},\ and\
  \citenamefont {Kwon}}]{Kang41}%
  \BibitemOpen
  \bibfield  {author} {\bibinfo {author} {\bibfnamefont {M.-S.}\ \bibnamefont
  {Kang}}, \bibinfo {author} {\bibfnamefont {I.-J.}\ \bibnamefont {Roh}},
  \bibinfo {author} {\bibfnamefont {Y.~G.}\ \bibnamefont {Lee}}, \bibinfo
  {author} {\bibfnamefont {S.-H.}\ \bibnamefont {Baek}}, \bibinfo {author}
  {\bibfnamefont {S.~K.}\ \bibnamefont {Kim}}, \bibinfo {author} {\bibfnamefont
  {B.-K.}\ \bibnamefont {Ju}}, \bibinfo {author} {\bibfnamefont {D.-B.}\
  \bibnamefont {Hyun}}, \bibinfo {author} {\bibfnamefont {J.-S.}\ \bibnamefont
  {Kim}}, \ and\ \bibinfo {author} {\bibfnamefont {B.}~\bibnamefont {Kwon}},\
  }\href@noop {} {\bibfield  {journal} {\bibinfo  {journal} {Sci. Rep.}\
  }\textbf {\bibinfo {volume} {6}},\ \bibinfo {pages} {26507} (\bibinfo {year}
  {2016})}\BibitemShut {NoStop}%
\bibitem [{\citenamefont {Roh}\ \emph {et~al.}(2016)\citenamefont {Roh},
  \citenamefont {Lee}, \citenamefont {Kang}, \citenamefont {Lee}, \citenamefont
  {Baek}, \citenamefont {Kim}, \citenamefont {Ju}, \citenamefont {Hyun},
  \citenamefont {Kim},\ and\ \citenamefont {Kwon}}]{Roh42}%
  \BibitemOpen
  \bibfield  {author} {\bibinfo {author} {\bibfnamefont {I.-J.}\ \bibnamefont
  {Roh}}, \bibinfo {author} {\bibfnamefont {Y.~G.}\ \bibnamefont {Lee}},
  \bibinfo {author} {\bibfnamefont {M.-S.}\ \bibnamefont {Kang}}, \bibinfo
  {author} {\bibfnamefont {J.-U.}\ \bibnamefont {Lee}}, \bibinfo {author}
  {\bibfnamefont {S.-H.}\ \bibnamefont {Baek}}, \bibinfo {author}
  {\bibfnamefont {S.~K.}\ \bibnamefont {Kim}}, \bibinfo {author} {\bibfnamefont
  {B.-K.}\ \bibnamefont {Ju}}, \bibinfo {author} {\bibfnamefont {D.-B.}\
  \bibnamefont {Hyun}}, \bibinfo {author} {\bibfnamefont {J.-S.}\ \bibnamefont
  {Kim}}, \ and\ \bibinfo {author} {\bibfnamefont {B.}~\bibnamefont {Kwon}},\
  }\href@noop {} {\bibfield  {journal} {\bibinfo  {journal} {Sci. Rep.}\
  }\textbf {\bibinfo {volume} {6}},\ \bibinfo {pages} {39131} (\bibinfo {year}
  {2016})}\BibitemShut {NoStop}%
\bibitem [{\citenamefont {Putilin}\ and\ \citenamefont
  {Yuragov}(2003)}]{Putilin43}%
  \BibitemOpen
  \bibfield  {author} {\bibinfo {author} {\bibfnamefont {A.~B.}\ \bibnamefont
  {Putilin}}\ and\ \bibinfo {author} {\bibfnamefont {E.~A.}\ \bibnamefont
  {Yuragov}},\ }\href@noop {} {\bibfield  {journal} {\bibinfo  {journal}
  {Measurement Techniques}\ }\textbf {\bibinfo {volume} {46}},\ \bibinfo
  {pages} {1173} (\bibinfo {year} {2003})}\BibitemShut {NoStop}%
\bibitem [{\citenamefont {Mccarty}\ \emph {et~al.}(2012)\citenamefont
  {Mccarty}, \citenamefont {Thompson}, \citenamefont {Sharp}, \citenamefont
  {Thompson},\ and\ \citenamefont {Bierschenk}}]{Maccarty44}%
  \BibitemOpen
  \bibfield  {author} {\bibinfo {author} {\bibfnamefont {R.}~\bibnamefont
  {Mccarty}}, \bibinfo {author} {\bibfnamefont {J.}~\bibnamefont {Thompson}},
  \bibinfo {author} {\bibfnamefont {J.}~\bibnamefont {Sharp}}, \bibinfo
  {author} {\bibfnamefont {A.}~\bibnamefont {Thompson}}, \ and\ \bibinfo
  {author} {\bibfnamefont {J.}~\bibnamefont {Bierschenk}},\ }\href@noop {}
  {\bibfield  {journal} {\bibinfo  {journal} {J. Electron. Mater.}\ }\textbf
  {\bibinfo {volume} {41}},\ \bibinfo {pages} {1274} (\bibinfo {year}
  {2012})}\BibitemShut {NoStop}%
\bibitem [{\citenamefont {Vasilevskiy}\ \emph {et~al.}(2014)\citenamefont
  {Vasilevskiy}, \citenamefont {Simard}, \citenamefont {Masut},\ and\
  \citenamefont {Turenne}}]{Vasilevskiy45}%
  \BibitemOpen
  \bibfield  {author} {\bibinfo {author} {\bibfnamefont {D.}~\bibnamefont
  {Vasilevskiy}}, \bibinfo {author} {\bibfnamefont {J.-M.}\ \bibnamefont
  {Simard}}, \bibinfo {author} {\bibfnamefont {R.~A.}\ \bibnamefont {Masut}}, \
  and\ \bibinfo {author} {\bibfnamefont {S.}~\bibnamefont {Turenne}},\
  }\href@noop {} {\bibfield  {journal} {\bibinfo  {journal} {J. Electron.
  Mater.}\ }\textbf {\bibinfo {volume} {44}},\ \bibinfo {pages} {1733}
  (\bibinfo {year} {2014})}\BibitemShut {NoStop}%
\bibitem [{\citenamefont {Beltran-Pitarch}\ \emph {et~al.}(2019)\citenamefont
  {Beltran-Pitarch}, \citenamefont {Prado-Gonjal}, \citenamefont {Powell},\
  and\ \citenamefont {Garcia-Canadas}}]{Pitarch36}%
  \BibitemOpen
  \bibfield  {author} {\bibinfo {author} {\bibfnamefont {B.}~\bibnamefont
  {Beltran-Pitarch}}, \bibinfo {author} {\bibfnamefont {J.}~\bibnamefont
  {Prado-Gonjal}}, \bibinfo {author} {\bibfnamefont {A.~V.}\ \bibnamefont
  {Powell}}, \ and\ \bibinfo {author} {\bibfnamefont {J.}~\bibnamefont
  {Garcia-Canadas}},\ }\href@noop {} {\bibfield  {journal} {\bibinfo  {journal}
  {J. Appl. Phys.}\ }\textbf {\bibinfo {volume} {125}},\ \bibinfo {pages}
  {025111} (\bibinfo {year} {2019})}\BibitemShut {NoStop}%
\bibitem [{\citenamefont {Kirby}\ and\ \citenamefont {Laubitz}(1973)}]{Kirby}%
  \BibitemOpen
  \bibfield  {author} {\bibinfo {author} {\bibfnamefont {C.~G.}\ \bibnamefont
  {Kirby}}\ and\ \bibinfo {author} {\bibfnamefont {M.~J.}\ \bibnamefont
  {Laubitz}},\ }\href@noop {} {\bibfield  {journal} {\bibinfo  {journal}
  {Metrologia}\ }\textbf {\bibinfo {volume} {9}},\ \bibinfo {pages} {103}
  (\bibinfo {year} {1973})}\BibitemShut {NoStop}%
\bibitem [{\citenamefont {Dunn}\ \emph {et~al.}(2019)\citenamefont {Dunn},
  \citenamefont {Daou},\ and\ \citenamefont {Atkinson}}]{DunnThomson}%
  \BibitemOpen
  \bibfield  {author} {\bibinfo {author} {\bibfnamefont {I.~H.}\ \bibnamefont
  {Dunn}}, \bibinfo {author} {\bibfnamefont {R.}~\bibnamefont {Daou}}, \ and\
  \bibinfo {author} {\bibfnamefont {C.}~\bibnamefont {Atkinson}},\ }\href@noop
  {} {\bibfield  {journal} {\bibinfo  {journal} {Rev.Sci. Instrum.}\ }\textbf
  {\bibinfo {volume} {90}},\ \bibinfo {pages} {024902} (\bibinfo {year}
  {2019})}\BibitemShut {NoStop}%
\bibitem [{\citenamefont {Dames}\ and\ \citenamefont
  {Chen}(2005)}]{Dames3omega}%
  \BibitemOpen
  \bibfield  {author} {\bibinfo {author} {\bibfnamefont {C.}~\bibnamefont
  {Dames}}\ and\ \bibinfo {author} {\bibfnamefont {G.}~\bibnamefont {Chen}},\
  }\href@noop {} {\bibfield  {journal} {\bibinfo  {journal} {Rev. Sci.
  Instrum.}\ }\textbf {\bibinfo {volume} {76}},\ \bibinfo {pages} {124902}
  (\bibinfo {year} {2005})}\BibitemShut {NoStop}%
\bibitem [{\citenamefont {Lu}\ \emph {et~al.}(2001)\citenamefont {Lu},
  \citenamefont {Yi},\ and\ \citenamefont {Zhang}}]{Lu3omega}%
  \BibitemOpen
  \bibfield  {author} {\bibinfo {author} {\bibfnamefont {L.}~\bibnamefont
  {Lu}}, \bibinfo {author} {\bibfnamefont {W.}~\bibnamefont {Yi}}, \ and\
  \bibinfo {author} {\bibfnamefont {D.~L.}\ \bibnamefont {Zhang}},\ }\href@noop
  {} {\bibfield  {journal} {\bibinfo  {journal} {Rev. Sci. Instrum.}\ }\textbf
  {\bibinfo {volume} {72}},\ \bibinfo {pages} {2996} (\bibinfo {year}
  {2001})}\BibitemShut {NoStop}%
\end{thebibliography}%
\end{document}